%% file: main.tex
\documentclass[conference]{Preamble/IEEEtran}
\pdfoutput=1

\input{Preamble/commands}
\input{Preamble/preamble}
\input{Preamble/macros}

\renewcommand\IEEEkeywordsname{Keywords}

\begin{document}

\pagestyle{plain} 

\title{\THEMIS: A Decentralized Privacy-Preserving\\ Ad Platform with Reporting Integrity}

\author{
	{\rm Gonçalo Pestana}\\ 
	Brave Software \and
	{\rm I\~{n}igo Querejeta-Azurmendi}\\
	Brave Software,\\UC3M
	\and
	{\rm Panagiotis Papadopoulos}\\ 
	Brave Software
	\and
	{\rm Benjamin Livshits}\\ 
	Brave Software,\\Imperial College London
}

\renewcommand{\emph}[1]{\textit{#1}}

\maketitle

\input{Sections/00_abstract}

\input{Sections/01_introduction}
\input{Sections/02_background}
\input{Sections/03_requirements}
\input{Sections/04a_design}
\input{Sections/04b_themis}
\input{Sections/06_evaluation}
\input{Sections/07_discussion}

\input{Sections/08_related}
\input{Sections/09_conclusions}

\bibliographystyle{unsrt}
\bibliography{main}

\appendix
\input{Sections/10_appendix_smart_contract}
\balance
\end{document}

%% file: Preamble/preamble.tex
\usepackage{tablefootnote}
\usepackage{times}  
\usepackage{thmtools}
\usepackage{thm-restate}
\usepackage{etoolbox}
\usepackage[dvipsnames]{xcolor}
\usepackage{bigstrut}
\usepackage{booktabs}
\usepackage{balance}
\usepackage{graphicx}
\usepackage{caption}
\usepackage[T1]{fontenc}
\usepackage[utf8]{inputenc}
\usepackage{lmodern}  
\usepackage{listings}
\usepackage{xspace}
\usepackage{amsmath}
\usepackage{amsthm}

\theoremstyle{definition}
\newtheorem{phase}{Phase}[subsection]

\usepackage{tikz}
\usepackage[noadjust]{cite}
\usepackage[hyphens]{url}
\usepackage{enumerate}
\usepackage[hidelinks]{hyperref}
\hypersetup{breaklinks=true}
\usepackage{pgfplots} 
\usepackage{pgfplotstable}
\usepackage{ulem} 

\pgfplotstableread{
x         y    y-max  y-min
Request  1.256 0.02 0.025
Fetch/Decrypt/Recover 0.469 0.015 0.011 
}{\client}

%% file: Preamble/macros.tex
\newcommand{\code}[1]{\textsf{#1}\xspace}

\newcommand{\nrads}{n_A}

\newcommand{\advs}{$Advs$\xspace}
\newcommand{\cp}{\texttt{Con.Part}\xspace}
\newcommand{\nrcp}{n} 
\newcommand{\threshold}{k}

\newcommand{\ad}{\code{Ad}}

\newcommand{\empirical}[1]{#1}

\newcommand{\ecenc}{\code{Encrypt}}
\newcommand{\ecdec}{\code{Decrypt}}

\newcommand{\pk}{\code{pk}}
\newcommand{\sk}{\code{sk}}

\newcommand{\pkbrave}{\pk_{\code{CM}}}
\newcommand{\skbrave}{\sk_{\code{CM}}}

\newcommand{\signsign}{\code{Sig.Sign}}
\newcommand{\signverify}{\code{Sig.Verify}}

\newcommand{\signature}{\sigma}

\newcommand{\distpk}{\pk_{\code{T}}}
\newcommand{\distsk}{\sk_{\code{T}}}

\newcommand{\distski}[1]{\sk_{\code{T},#1}}

\newcommand{\encryptedad}{\code{Enc}.\ad}

\newcommand{\encryptedaggregatedresult}{\code{Aggr.Res}}
\newcommand{\encryptedaggregatedclicks}{\code{Aggr.Clicks}}
\newcommand{\decryptedaggr}{\code{Dec.\encryptedaggregatedresult}}
\newcommand{\result}{\code{Res}}
\newcommand{\aggrproofdec}{\Pi_{\result}}
\newcommand{\aggrclicks}{\code{Aggr.Clicks}}

\newcommand{\vrf}{\code{VRF}}
\newcommand{\vrfpk}{\vrf.\pk}
\newcommand{\vrfsk}{\vrf.\sk}

\newcommand{\vrfrandgen}{\vrf.\code{RandGen}}
\newcommand{\vrfrandom}{\vrf.\code{Rand}}
\newcommand{\vrfproof}{\Pi^{\vrf}}
\newcommand{\vrfseed}{\epsilon}
\newcommand{\vrfmax}{\code{MAX.DRAW}}

\newcommand{\encvec}{\code{EncVec}}
\newcommand{\encvecpublic}{\code{EncVec'}}

\newcommand{\signaturereward}{\code{SignReward}}

\newcommand{\policy}{\mathcal{P}}
\newcommand{\encpolicy}{\code{Enc }\mathcal{P}}
\newcommand{\seckey}{\mathcal{S}}
\newcommand{\encseckey}{\code{Enc }\mathcal{S}}
\newcommand{\valikey}{\pk_{\mathcal{V}}}

\newcommand{\addr}{\code{Addr}}
\newcommand{\payreq}{\mathcal{E}}

\newcommand{\cfpk}{\code{\cf.}\pk}

\newcommand{\adclicks}{\code{ac}}%
\newcommand{\nrad}{\code{n}}%

\newcommand{\THEMIS}{THEMIS\xspace}
\newcommand{\PoA}{\textsc{PoA}\xspace}
\newcommand{\PoW}{\textsc{PoW}\xspace}
\newcommand{\PSC}{\textsc{PSC}\xspace}
\newcommand{\FSC}{\textsc{FSC}\xspace}

\newcommand{\cf}{CF\xspace}
\newcommand{\cfs}{CFs\xspace}
\newcommand{\cm}{CM\xspace}
\newcommand{\sidechain}{sidechain\xspace}
\newcommand{\sidechains}{sidechains\xspace}

\newcommand{\eg}{{e.g.,}\xspace}
\newcommand{\etc}{{etc.}\xspace}

\newcommand{\ie}{{i.e.,}\xspace}

\newcommand{\enc}{\code{Enc}}
\newcommand{\dec}{\code{Dec}}
\newcommand{\sign}{\code{Sign}}
\newcommand{\dpoolsize}{L\xspace} 

\newcommand{\paymentsPerDay}{\empirical{1.7M}\xspace}
\newcommand{\batchsize}{\empirical{800}\xspace}
\newcommand{\paymentsPerMonth}{\empirical{50.9M}\xspace}
\newcommand{\usersScalability}{\empirical{51M}\xspace}
\newcommand{\usersScalabilityMulti}{\empirical{153M}\xspace}

\newcommand{\rewardrequest}{\empirical{4.69 seconds}\xspace}
\newcommand{\rewardverif}{\empirical{5.04 seconds}\xspace}

\newcommand{\partdec}{\code{Part.Dec}\xspace}
\newcommand{\decsharei}[1]{\code{Dec.Share}_{#1}}

%% file: Sections/00_abstract.tex
\begin{abstract}
Online advertising fuels the (seemingly) free internet. However, although users can access most websites free of charge, they need to pay a heavy cost on their privacy and blindly trust third parties and intermediaries that absorb great amounts of ad revenues and user data. This is one of the reasons users opt out from advertising by resorting ad blockers that in turn cost publishers millions of dollars in lost ad revenues.

Existing privacy-preserving advertising approaches (\eg Adnostic, Privad, Brave Ads) from both industry and academia cannot guarantee the integrity of the performance analytics they provide to advertisers, while they also rely on centralized management that users have to trust without being able to audit.

In this paper, we propose \THEMIS, a novel privacy-by-design ad platform that is decentralized and requires zero trust from users. \THEMIS (i) provides auditability to all participants, (ii) rewards users for viewing ads, and (iii) allows advertisers to verify the performance and billing reports of their ad campaigns. To demonstrate the feasibility and practicability of our approach, we implemented a prototype of \THEMIS using a combination of smart contracts and zero-knowledge schemes. Performance evaluation results show that during ad reward payouts, \THEMIS can support more than~\usersScalability users on a single-\sidechain setup or ~\usersScalabilityMulti users on a multi-\sidechain setup, thus proving that \THEMIS scales linearly. 
\end{abstract}

%% file: Sections/01_introduction.tex
\section{Introduction}
\label{sec:intro}

Digital advertising is the most popular way of funding websites, despite the many other alternative monetization systems that have been proposed~\cite{paywalls,wikipediaFunds1,truth2018,10.1145/1963405.1963451}.
However, web advertising has fundamental flaws, including market fragmentation~\cite{frag1, frag2}, rampant fraud~\cite{fraud1, fraud2,adfraud,adfraud2,Liu:2014:DDC:2616448.2616455,Zarras:2014:DAM:2663716.2663719,malvertising}, centralization around intermediaries (\eg Supply/Demand-side platforms, Ad exchanges, Data Management Platforms) who absorb half of spent advertising dollars~\cite{middlemenHalf} and unprecedented privacy harm from the extensive tracking of users ~\cite{trackersWWW2016,razaghpanah2018apps,exclusiveCSync,englehardt2016online}. Web advertising is increasingly being ignored or blocked by users, too.
The average click-through-rate today is~2\% ~\cite{ctrOverall}, while at the same time~47\% of Internet users globally use ad-blockers~\cite{adblockreport2019}, thus costing publishers millions of dollars in ad revenues every year~\cite{adblockersCost}

\point{Alternative advertising strategies}
Academia and industry have responded to some of these challenges by designing new monetization approaches. These systems generally emphasize end-user choice, privacy protections, fraud prevention, or performance improvements. Brave Ads~\cite{Brave2017}, Privad~\cite{privad}, and Adnostic~\cite{toubiana2010adnostic} are among the three most prominent such proposals. These systems, while promising, have limitations that prevent them from being compelling replacements for the existing web advertising system. Specifically, these systems either do not scale well, or require the user to trust central authorities within the system to process ad transactions. Additionally, these proposed systems lack auditability: users need to blindly trust the ad network that exclusively determines how much advertisers will be charged, as well as the revenue share the publishers will get~\cite{7164916}. Malicious ad networks can overcharge advertisers or underpay publishers, while malicious advertisers can deny actual views/clicks and ask for refunds, since bills frequently cannot be adequately justified by the ad network (non-repudiation). 

\begin{figure}[bt]
    \centering
    \includegraphics[width=.95\columnwidth]{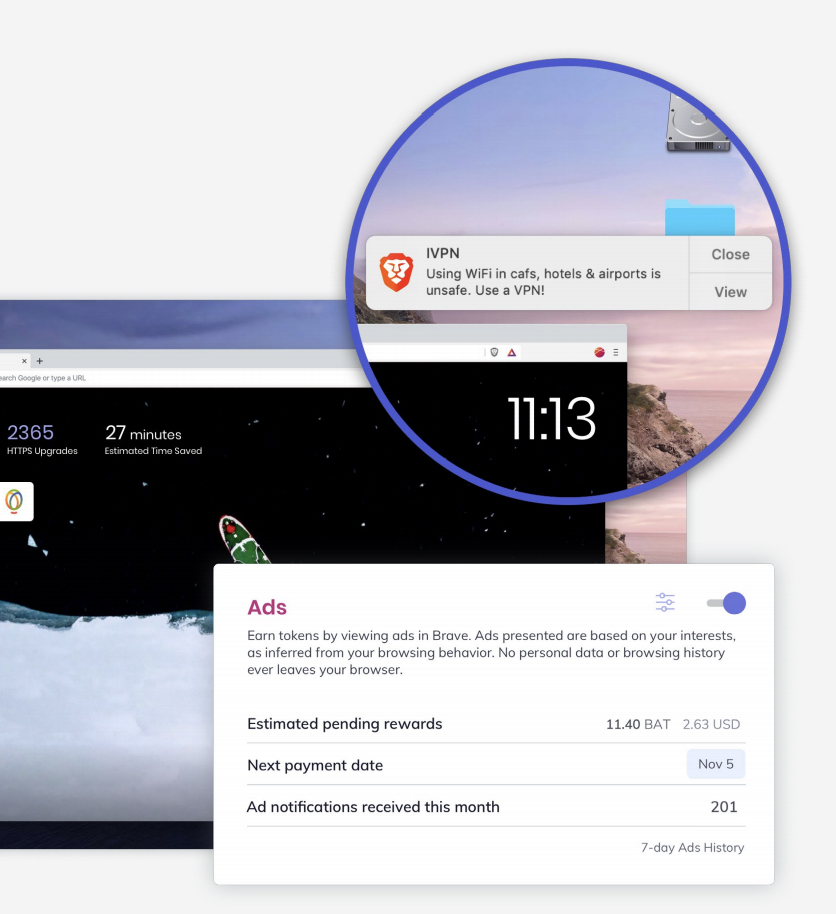}
    \caption{Brave Rewards in action.}
    \label{fig:rewards}
\end{figure}

\point{Compensating ad viewing}
The idea of compensating end-users to view ads is not so far-fetched: \emph{Brave Ads}~\cite{Brave2017}, in operation since~2019, is a browser-based advertising system that 
compensates users for their \emph{attention} while viewing ads. As can be seen in Figure~\ref{fig:rewards}, a Brave browser user is shown an ad as an operating system notification, for which they are compensated in the form of a cryptocurrency in their built-in browser wallet.


\point{Introducing \THEMIS}
To address these issues, in this paper, we propose \THEMIS, the first decentralized, scalable, and privacy-preserving ad platform that provides auditability to its participants. Thus, users do not need to trust any protocol actor (\eg to protect their privacy or handle payments). To increase the end-user engagement with ads and provide the advertisers with feedback regarding their campaigns, \THEMIS (i) incentivizes users to interact with ads by rewarding them for their ad viewing and (ii) provides advertisers with verifiable and privacy-preserving campaign performance analytics. This way, advertisers can accurately learn \emph{how many users viewed their ads, but without learning which users}. 

To achieve the above, our system uses a \sidechain design pattern and smart contracts to eliminate the centralized management and brokerage of the current ad ecosystem. To reward users for their ad viewing and perform reward payouts without revealing user interests, \THEMIS leverages a partial homomorphic encryption scheme. According to existing proposals~\cite{ctrBrave} paying users for viewing ads successfully motivates them to click on these ads (14\% click-through-rate compared to the~2\% of traditional ad ecosystem). \THEMIS is privacy-preserving; it sends zero bits of behavioral user data outside of the user device, while at the same time, it serves ads relevant to user's interests by matching these ads locally to the their profile. \THEMIS leverages zero-knowledge schemes during ad operations, so every actor in the system can cryptographically verify at any time that everybody follows the protocol as expected. Finally, by using a multiparty computation protocol, \THEMIS guarantees the computational integrity of the campaign performance analytics it provides to advertisers.


\point{Contributions}
In summary, we make the following contributions to the goal of a decentralized, privacy-respecting advertising system:
\begin {enumerate}
\item We propose \THEMIS, a novel privacy-preserving advertising platform that rewards users for viewing ads. Contrary to existing proposals, our system is decentralized and leverages smart contracts to orchestrate reward calculation and payments between users and advertisers. This way, our platform avoids relying on a single trusted central authority.

\item To assess the feasibility of our approach, we implement a prototype of our system in Rust and Solidity. 
We provide the source code of \THEMIS publicly~\cite{themis-repo}. 

\item To evaluate the scalability and practicability of our system, we perform experiments on a single- and different multi-\sidechain setups. We see that \THEMIS can support reward payments of around~\usersScalability  users per month on a single \sidechain or~\usersScalabilityMulti users on a parallel setup of three \sidechains, thus proving that \THEMIS scales linearly.
\end {enumerate}

%% file: Sections/02_background.tex
\section{Building Blocks}
\label{sec:blocks}

In this section, we introduce the techniques and mechanisms leveraged throughout the paper and describe why and how \THEMIS uses them.

\subsection{Proof of Authority Blockchains}
\label{subsec:poa}
\THEMIS relies on a blockchain with smart contract functionality to provide a decentralized ad platform. Smart contracts enable us to perform all payments without trusting a single central authority. Ethereum Mainnet~\cite{eth} is a such popular smart contract-based blockchain. However, due to its low transaction throughput, the high gas costs, and the overall poor scalability, in \THEMIS, we chose to use a \mbox{Proof-of-Authority}~(\PoA) blockchain instead. 

Consensus protocols constitute the basis of any distributed system. The decision on which consensus mechanism to use affects the properties, scalability and assumptions of the services build on top of the distributed system~\cite{consensus_sok}. A \PoA blockchain consists of a distributed ledger that relies on consensus achieved by a permissioned pool of validator nodes. \PoA validators can rely on fast consensus protocols such as IBFT/IBFT2.0~\cite{EIP650, ibft} and Clique~\cite{clique}, which result in faster minted blocks and thus \PoA can reach higher transaction throughput than traditional \PoW based blockchains. As opposed to traditional, permissionless blockchains (such as Bitcoin~\cite{btc} and Ethereum~\cite{eth}), the number of nodes participating in the consensus is relatively small and all nodes are authenticated. In \THEMIS, the set of validators may consist of publishers or foundations.


\point{Private input transactions}
\label{sec:private_transactions} \THEMIS leverages private input transactions to maintain the privacy of the advertisers. More precisely, \THEMIS uses the private input transactions as defined by the Quorum \PoA \sidechain \cite{quorumpoa}. Providing private input functionality in smart contracts requires the inputs to be encrypted with all validator's public keys. 
By encrypting both inputs and outputs with validator's public keys, the parameters are private from readers of the public information while, at the same time, validator nodes can decrypt the values and run the smart contracts correctly in order to achieve consensus. For simplicity, we refer to the public keys of validators as a single one throughout the paper, and denote it with $\valikey$. Projects like Quorum~\cite{quorumpoa, quorumgo} and Hyperledger Besu~\cite{pegasys} are \mbox{Ethereum-based} distributed ledger implementations that implement private transaction inputs and outputs.

\subsection{Cryptographic Tools}
\label{sec:cryptoTools}

\point {Confidentiality}
\THEMIS uses an additively homomorphic encryption scheme to calculate the reward payouts for each user, while keeping the user behavior (e.g. ad clicks) private. Given a \mbox{public-private} \mbox{key-pair} $(\pk, \sk)$, the encryption scheme is defined by three functions: first, the encrypt function, where given a public key and a message, it outputs a ciphertext, $C = \enc(\pk, M)$. Secondly, the decrypt function, that given a ciphertext and a private key, it outputs a decrypted message, $M = \dec(\sk, C)$. And finally, the signing function, where given a message and a secret key, it outputs a signature on the message, $S = \sign(\sk, M)$. The additive homomorphic property guarantees that the addition of two ciphertexts, $C_1 = \enc(\pk, M_1), C_2 = \enc(\pk, M_2)$ encrypted under the same key, results in the addition of the encryption of its messages, more precisely, 
\begin{equation*}
    C_1 + C_2 = \enc(\pk, M_1) + \enc(\pk, M_2) = \enc(\pk, M_1 + M_2).
\end{equation*}
Some examples of such encryption algorithms are ElGamal \cite{elgamalcrypto} or Paillier \cite{paillier} encryption schemes. We run our experiments using ElGamal over elliptic curves. 

\point{Integrity} 
To prove correct decryption, \THEMIS leverages zero knowledge proofs (ZPKs)~\cite{Goldwasser:1985:KCI:22145.22178} which allow an entity (the prover) to prove to a different entity (the verifier) that a certain statement is true over a private input. This proof does not disclose any other information from the input other than whether the statement is true or not. \THEMIS leverages ZKPs to offload computation on the client-side, while maintaining integrity and privacy. More precisely, a user decrypts the earned reward payout, and proves that the decryption is correct. We denote proofs with the letter $\Pi$, and use $\Pi.\code{Verify}$ to denote verification of a proof. 

\point{Distribution of trust}
\THEMIS distributes trust to generate a \mbox{public-private} \mbox{key-pair} for each ad campaign, under which sensitive user information is encrypted. For that end, \THEMIS leverages a distributed key generation~(DKG) protocol. This allows a group of participants to distributively generate the \mbox{key-pair} $(\distpk, \distsk)$, where each participants has a share of the private key, $\distski i$, and no participant ever gains knowledge of the full private key, $\distsk$.
The resulting \mbox{key-pair} is a $\threshold - \nrcp$ threshold \mbox{key-pair}, which requires at least $\threshold$ out of the $\nrcp$ participants that distributively generated the key, to interact during the decryption or signing protocols. There exists both synchronous~\cite{Gennaro2007} and asynchronous~\cite{asyncdkg1, asyncdkg2} schemes. In our scenario, building on top of a synchronous scheme is acceptable, as the number of participants is small, and participants are incentivised to be online during the key generation procedure. Hence, we follow the protocol presented in~\cite{Gennaro2007} for the key generation as well as the decryption procedure. 

In order to chose this selected group of key generation participants in a distributed way, \THEMIS leverages verifiable random functions~(VRFs) \cite{micalivrf, irtf-cfrg-vrf-05}. In general, VRFs enable users to generate a random number and prove its randomness. In \THEMIS, we use VRFs to select a random pool of users and generate the distributed keys. Given a \mbox{public-private} \mbox{key-pair} $(\vrfpk, \vrfsk)$, VRFs are defined by one function: random number generation, which outputs a random number, \vrfrandom and a zero knowledge proof of correct generation, $\vrfproof$, 
\begin{equation*}
    (\vrfrandom, \vrfproof) = \vrfrandgen(\vrfsk, \vrfseed),
\end{equation*}
where $\vrfseed$ is a random seed.

\point{Confidential payments for \mbox{account-based} blockchains} 
\label{sec:conf_payments_poa}
Confidential payments on \mbox{account-based} blockchains allow transfers of assets between accounts without disclosing the amount of assets being transferred or the balance of the accounts. Additionally, the sender can prove the correctness of the payment (\ie prove that there was no double spending). Confidential payments have drawn a lot of interest in both academia~\cite{aztecpaper,zether, zcashpaper} and industry~\cite{zcash, monero, anonymouszether} recently. We studied the throughput of two solutions, namely AZTEC~\cite{aztecpaper} and the Zether~\cite{zether} protocols for \THEMIS.

The AZTEC protocol implements a toolkit and a set of smart contracts for building confidential assets on top of the Ethereum virtual machine~\cite{evm}. The AZTEC protocol defines a commitment scheme and zero-knowledge proofs for verifying and validating transactions without disclosing the balance of the asset transaction. An important feature of AZTEC is that it enables the prover to generate proof of correct payments in batches, which amortizes the costs of multiple payments.

The Zether protocol uses \mbox{Sigma-Bulletproofs} and one-out-of-many proofs to achieve confidential and anonymous payments in \mbox{account-based} blockchains. Zether does not provide batching of payment proof validations, which means that the time for settling payments grows linearly with the number of payments issued. We this reason we selected AZTEC as our underlying private payment system. In Section~\ref{sec:results} we show how AZTEC payments allow \THEMIS to scale up to \paymentsPerMonth payments per month.  

%% file: Sections/03_requirements.tex
\section{Threat Model and Goals}
\label{sec:motivation}
In this section, we introduce the participants of \THEMIS and the threat model of the system. In addition, we describe the design principles of \THEMIS and how existing systems compare with it.

\subsection{Main Actors}
\point{\PoA validator nodes} \THEMIS leverages a \PoA \sidechain that relies on consensus achieved by a pool of permissioned validator nodes. The role of the validators is to mine the blocks of the \sidechain. In order to achieve this, each validator needs to evaluate the smart contract instructions against the user's inputs and global state, achieve consensus among the consortium on what is the next stage of the blockchain and mine new blocks. To preserve independence and the zero-trust requirement of the \sidechain, in \THEMIS, validator nodes are maintained by  non-colluding, independent third parties (\eg the Electronic Frontier Foundation (EFF)~\cite{eff}, or a non-profit trade foundation of applications),  similar to existing volunteer networks like Tor~\cite{dingledine2004tor}, Gnutella~\cite{gnutella} or distributed VPNs~\cite{hola,varvello2019vpn0}. 

\point{Campaign facilitator (\cf)} 
The \cf is an entity authorized by the \PoA consortium that helps running the \THEMIS protocol. A \cf interacts with advertisers to agree on an ad policy of their preference and deploys the smart contracts in the \PoA \sidechain. In addition, the \cf is responsible for performing the confidential and verifiable payments to the users. The \cf has the role of a facilitator; \THEMIS requires an honest run of the protocol by the \cf for completeness, but not for verifiability. The system can detect when the \cf misbehaves.

\point{Advertisers} 
The advertisers agree with the \cf the policies for each ad campaign they want to launch in the context of \THEMIS. They receive an anonymized feedback for the performance of their campaigns. Advertisers can verify the validity of the reporting. In addition, advertisers can interact with the \PoA chain to verify that the amount charged for running campaigns corresponds to valid user interactions with campaign ads.

\point{Users} 
The users interact with the ads through an advertising platform, which selects and distributes the ads. Users interact with the \PoA \sidechain so their rewards are computed and paid. \THEMIS users may also participate in a consensus pool where they interact with other users in a peer-to-peer way. We refer to them as consensus participants~(\cp).

\subsection{Goals and Comparison with Alternatives}
\label{subsec:reqs}

The goal of this paper is to design (i) a decentralized and (ii) trustless ad platform that (iii) is private-by-design while, at the same time, (iv) rewards users for the attention they give while viewing ads and (v) provides metrics for the ad campaigns of the advertisers. The key system properties we focus on while designing \THEMIS, include privacy (for both user interactions and advertiser policies), decentralization and auditability (by providing verifiable rewards billing and campaign reporting integrity) and scalability:

\begin{enumerate}
\item \textbf{Privacy.} In the context of a sustainable ad ecosystem, we define privacy as the ability for not only users, but also advertisers to use our system without disclosing any critical information:
\begin{enumerate}
\item For the user, privacy means being able to interact (\ie view, click) with ads without revealing their interests/preferences to any third party. 
    In \THEMIS, we preserve the privacy of the user not only when they are interacting with ads but also when they claim the corresponding rewards for these ads. 

\item For the advertisers, privacy means that they are able to setup ad campaigns without revealing any policies (\ie what is the reward of each of their ads) to the prying eyes of their competitors. To achieve that, \THEMIS keeps these ad policies confidential throughout the whole process, while it enables users to provably claim rewards based on the ad policies.
\end{enumerate}

\item \textbf{Decentralization and auditability.}
Existing ad platforms~\cite{Brave2017,privad,toubiana2010adnostic} require a single central authority to manage and orchestrate the execution of their protocols. Both privacy and billing is dependent on the correct behaviour of the single authorities. 

However, as nicely pointed out in~\cite{7164916}, what if this --- considered as \emph{trusted}~---~entity censors users by denying or transferring incorrect amount of rewards? What if it attempts to charge advertisers more than what they should pay based on users' ad interactions? What if the advertising policies are not applied as agreed with the advertisers when setting up ad campaigns?

One of the primary goals of our system is to be decentralized and require no trust from users. To achieve this, \THEMIS leverages a Proof-of-Authority (PoA) blockchain with smart contract functionality. To provide auditability, \THEMIS leverages zero-knowledge proofs to ensure the correctness and validity of both billing and reporting thus allowing all actors to verify the authenticity of the statements and the performed operations.

\item \textbf{Scalability.}
Ad platforms need to be able to scale seamlessly and serve millions of users. However, important proposed systems fail to achieve this~\cite{privad,toubiana2010adnostic}. In this paper, we consider scalability an important aspect affecting the practicability of a system. \THEMIS needs to not only serve ads in a privacy preserving way to millions of users but also finalize the payments related to their ad rewards in a timely and resource efficient manner.

\end{enumerate}

\subsection{Threat Model}
\label{subsec:threat model}
In \THEMIS, we assume computationally bounded adversaries capable of (i) snooping communications, (ii) performing replay attacks, or (iii) cheating by not following the protocol.

One such adversary may act as a \cf aiming to collect more processing fees than agreed at the cost of user rewards or advertiser refunds. 
Another adversary may attempt to breach the user privacy and snoop their ad interactions. Such information could reveal interests, political/sexual/religious preferences,  that can be later sold or used beyond the control of the user~\cite{radioshakData,brokserSell,toysmartData,Gill:2013:BPF:2504730.2504768}. Such an adversary may control at most $\threshold$ of the $\nrcp$ (see Section~\ref{sec:cryptoTools}) randomly selected users that are part of the consensus pool. This adversary is an \emph{adaptive} adversary, meaning that it can decide which participants to corrupt based on prior observations (\eg once the consensus pool has been selected). We require that $\threshold < \nrcp / 2$, which is optimal in the threshold encryption scenario~\cite{Gennaro2007}. Given that the consensus participants are chosen randomly, this means that, for ad interaction privacy, we assume that at most half of the consensus participants are malicious. 
Other adversaries may try to break the confidentiality of the advertisers' agreed ad policies and disclose rewarding strategies to competitors. We assume that such an adversary may act as a user and/or advertiser in the protocol, but cannot control the campaign facilitators or the \PoA validators. 
%
%

\point{Out-of-scope Attacks} 
We acknowledge that client-side fraud, together with malvertising and brand safety are  important issues of the ad industry. However, similar to the related work~\cite{toubiana2010adnostic,10.5555/646139.680791,10.1145/2462456.2464436,privad,green2016protocol}, in this paper, we do not claim to address all issues of digital advertising. 
There is an abundance of papers aiming to detect and prevent cases of client-side fraud (\ie bot clicks, click farms, sybil attacks), which can be also used in the context of THEMIS (\eg distributed user reputation systems~\cite{yang2019decentralized}, anomaly detection, bluff ads~\cite{haddadi2010fighting}, bio-metric systems~\cite{2019zksense}, client puzzles~\cite{4215910}, \etc).


%% file: Sections/04a_design.tex
\section{System Overview}
\label{sec:system}
In this section, we describe in detail the \THEMIS system. We begin with a straw-man approach to describe the basic principles of the system. We build on this straw-man approach and, step-by-step, we introduce the decentralized and trustless ad platform. For presentation purposes, in the rest of this paper, we assume that users interact with \THEMIS through a web browser, although users may interact with it through a mobile app in the exact same way. For the construction of \THEMIS, we assume the existence of privacy-preserving ad personalization and incentives for ad-viewing in the advertising platform. These are two techniques that have been successful in industry products. 

\subsubsection{Privacy-preserving Ad Personalization}
To perform privacy-preserving ad personalization, \THEMIS adheres to the paradigm of Adnostic~\cite{toubiana2010adnostic} and of Brave Ads~\cite{Brave2017}, which has been in continuous operation since 2019~\cite{BraveAds}. Similarly to these systems, users periodically download an \emph{ad catalog} from the ad server (maintained by the \cm). The ad catalog includes data and metadata for ads from all active campaigns. 

The ad-matching happens locally based on a pre-trained model and the interests of the user are extracted their web browsing history, in a similar way as in~\cite{Brave2017, privad}. No data leaves the user's device, thus creating a walled garden of browsing data that are used for recommending the best-matching ad while user privacy is guaranteed.

The idea of prefetching ads is not new~\cite{levin2009nurikabe,10.1145/2462456.2464436,privad,toubiana2010adnostic,BraveAds}. Studies have shown that pre-fetching ads in bulk for smartphone users is not only practical but can reduce energy consumed during ad transactions by more than~50\%~\cite{mohan2013prefetching}.

\subsubsection{Incentives for Ad-viewing}
\label{sec:straw_rewards}
\THEMIS, compensates users for the attention they pay to ad impressions, thus incentivizing them to interact with ads. Other academic and industry products use a similar user rewarding scheme~\cite{wang2015privacy,parra2017pay,Brave2017}. Rewarding schemes in production have shown an increase in the user engagement, providing high click-through-rates (\ie 14\% on average~\cite{ctrBrave}) on the ads shown to users.

In \THEMIS, each viewed/clicked ad yields a reward, which can be fiat money, crypto-coins, coupons, \etc. Different ads may provide different amount of reward to the users. This amount is agreed by the corresponding ad creator (\ie the advertiser) and the \cm.
Users \emph{claim} the rewards periodically (\eg every 2 days, every week or every month). The users must requests their reward for the ads they viewed and interacted with. 

\begin{figure}[tb]
    \centering
    \scriptsize
     \begin{tabular}{ll}
        \toprule
        \bf Notation & \bf Description \\
        \midrule
        $\encvec$ & Encrypted vector of ad clicks  \\
        \midrule
        $\encryptedaggregatedresult$ & Encrypted result of the aggregate calculated \\
        & over the $\encvec$ vector \\
        \midrule
        $\decryptedaggr$ & Result of the decryption of $\encryptedaggregatedresult$ by the \\ 
        & user  \\
        \midrule
        $\signaturereward$ & Signature of the aggregate computation that \\ 
        & is used by TA to verify that the $\decryptedaggr$ \\
        & is correct \\
        \midrule
        $\aggrproofdec$ & Proof of correct decryption of $\encryptedaggregatedresult$ \\
        \midrule
        $\vrfpk, \vrfsk$ & Ephemeral public-private key pair for players \\ 
        & to participate in the draw \\
        \midrule
        $\vrfseed$ & Random seed to generate random numbers \\
        \midrule
        $\vrfmax$ & Boundary for selecting consensus participants\\
        \midrule
        $\vrfrandom$ & Random number \\
        \midrule
        $\vrfproof$ & Proof of randomness based on VRF \\
        \midrule
        $\distpk$ & Threshold public key generated in a distributed \\ 
        & way by the consensus participants \\
        \midrule
        $\distski i$ & Share $i$ of the threshold private key \\
        \midrule
        $\decsharei i$ & Result of partial decryption using $\distski i$ \\
        \midrule
        $\encryptedaggregatedclicks$ & Value of the aggregate clicks by all users\\
        \bottomrule
     \end{tabular}
    \caption{Summary of our cryptographic notation.}
    \label{tbl:notation}
\end{figure}

\subsection{A Straw-man Approach}
\label{sec:strawman}
Our straw-man approach is the first step towards a privacy-preserving and trustless online advertising system. Our goal at this stage is to provide a mechanism for advertisers to create ad campaigns and to be correctly charged when their respective ads are delivered to users. In addition, the system aims at keeping track of the ads viewed by users, so that (i) advertisers can have feedback about the performance of the ad campaigns and (ii) users can be rewarded for interacting with ads. All these goals should be achieved while preserving ad policy privacy and user privacy.

We assume three different roles in the straw-man version of \THEMIS: (i) the users, (ii) the advertisers, and (iii) an ad campaigns manager (\cm). The users are incentivized to view and interact with ads created by the advertisers. The \cm is responsible (a) for orchestrating the protocol, (b) for handling the ad views reporting and finally (c) for calculating the rewards that need to be paid to users according to the policies defined by the advertisers. Figure~\ref{tbl:notation} summarizes the notation used throughout this section.

Note that the straw-man version of \THEMIS relies on an ad campaign manager, which is a single central authority required for orchestrating the protocol. In addition, users and advertisers must trust the \cm. We use this simplified version to lay out the fundamental principles of the protocol. We present the improved -- and decentralized -- version of the protocol in Section~\ref{sec:demis}.

\begin{figure}[t]
    \centering
    \includegraphics[width=1.07\columnwidth]{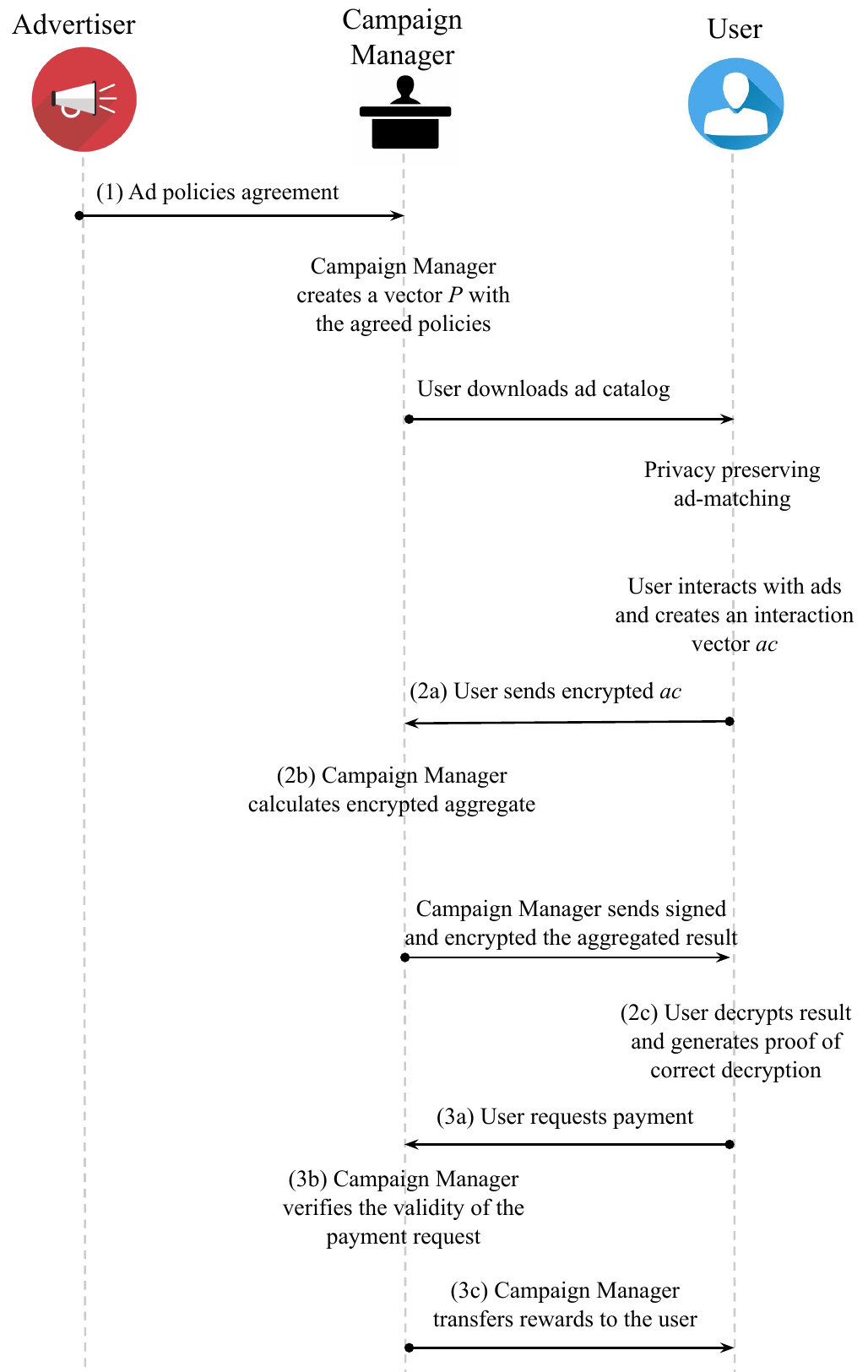}
    \caption{High-level overview of the user rewards claiming procedure of our straw-man approach. Advertisers can set how much they reward each ad click without disclosing that to competitors. The user can claim rewards without exposing which ads they interacted with.}
    \label{fig:diagram-strawman}
\end{figure}

In Figure~\ref{fig:diagram-strawman}, we present an overview of the reward and claiming procedure of the \THEMIS straw-man. 

\begin{phase}{\textbf{Defining Ad Rewards:}}
\label{phase:def-ads-strawman}
In order for an advertiser to include their ads in the ad campaign running on \THEMIS, they first need to agree with the \cm on the policies of the given campaign. An ad policy consists of the rewards a user should earn per ad visualization and engagement (step~1 in Figure~\ref{fig:diagram-strawman}). 

The \cm encodes the ad policies from multiple advertisers as a vector $\policy$, where each index corresponds to the amount of tokens that an ad yields when viewed/clicked (\eg Ad 1: 4 coins, Ad 2: 20 coins, Ad 3: 12 coins). The \cm stores this vector privately and the advertiser needs to trust that the policies are respected. Note, as referred earlier, the need for trust will be removed in the final version of \THEMIS (Section~\ref{sec:demis}). The indices used in $\policy$ are aligned with the ones of the ad catalog.

For the sake of simplicity, throughout this section, we consider one advertiser who participates in our ad platform and runs multiple ad campaigns. Of course, in a real world scenario many advertisers can participate and run many ad campaigns simultaneously. We also consider ``agreed policies'' as the amount of ``coins'' an ad provides as reward for a click by a user.
\end{phase}
\begin{phase}{\textbf{Claiming Ad Rewards:}}
\label{phase:claim-rewards-strawman}
The user locally creates an \emph{interaction vector}, which encodes information about the number of times each ad of the catalog was viewed/clicked (\eg Ad 1: was viewed 3 times, Ad 2: was viewed 0 times, Ad 3: was viewed 2 times). 

In every payout period, the user encrypts the state of the interaction vector. More specifically, let 
\begin{equation*}
\adclicks = \left[\nrad_1, \ldots, \nrad_{\nrads} \right]
\end{equation*}
be the interaction vector containing the number of views/clicks of users with each ad, where element $i$ of vector \adclicks represents the number of times ad $i$ was viewed/clicked. On every payout period, the user generates a new ephemeral key pair $(\pk, \sk)$, to ensure the unlinkability of the payout requests. By using this key, they encrypt each entry of \adclicks:
\begin{equation}
   \encvec = \left[\ecenc(\pk, \nrad_1), \ldots, \ecenc(\pk, \nrad_{\nrads}) \right]
   \label{eq:ac}
\end{equation}
and send \encvec to the \cm (step 2a in Figure~\ref{fig:diagram-strawman}).
\cm cannot decrypt the received vector and thus cannot learn the user's ad interactions (and consequently their interests). Instead, they leverage the additive homomorphic property of the underlying encryption scheme (as described in Section~\ref{sec:cryptoTools}) to calculate the sum of all payouts based on the interactions encoded in \encvec (step 2b in Figure~\ref{fig:diagram-strawman}). More formally, the \cm computes the aggregate payout for the user as follows:
\begin{equation*}
    \encryptedaggregatedresult = \sum_{i=1}^{N} \policy\left[i\right] \cdot \encvec\left[i\right],
    \label{eq:aggr_payout}
\end{equation*}
where $\policy_i$ is the ad policy associated with the ad in the position $i$ of the vector. Then \cm signs the computed aggregate result: 
\begin{equation*}
    \signaturereward =\signsign(\encryptedaggregatedresult, \skbrave)
\end{equation*}
and sends the 2-tuple $(\encryptedaggregatedresult, \signaturereward)$ back to the user.
Upon receiving this tuple (step 2c in Figure~\ref{fig:diagram-strawman}), the user verifies the signature of the result. If 
\begin{equation*}
\signverify(\encryptedaggregatedresult, \signaturereward) = \bot,
\end{equation*}
the user repeats the request to the \cm. If the signature is valid, the user proceeds with decrypting the result, 
\begin{equation*}
\decryptedaggr = \ecdec(\sk, \encryptedaggregatedresult).
\end{equation*} 
As a final step, it proves the correctness of the performed decryption by creating a zero knowledge proof of correct decryption, $\aggrproofdec$. 
\end{phase}

\begin{phase}{\textbf{Payment Request:}}
\label{phase:payment-request-strawman}
Finally, the user generates the payment request and sends the following 4-tuple to the \cm (step 3a in Figure~\ref{fig:diagram-strawman}):
\begin{equation*}
    \left(\decryptedaggr, \encryptedaggregatedresult, \signaturereward, \aggrproofdec \right)
\end{equation*}
As a next step (step 3b in Figure~\ref{fig:diagram-strawman}), the \cm verifies that the payment request is valid. More specifically, the \cm will reject the payment request of the user if
\begin{equation*}
    \signverify(\pkbrave,\signaturereward, \encryptedaggregatedresult) = \bot 
\end{equation*}
or
\begin{equation*}
\texttt{Verify}(\aggrproofdec) = \bot
\end{equation*}
Otherwise, it proceeds with transferring the proper amount (equal to $\decryptedaggr$) of rewards to the user. 
\end{phase}

\point{In summary} 
As described, the straw-man \THEMIS guarantees that:
\begin{enumerate}[A.]
    \item The user receives the rewards they earned by interacting with ads. This happens without requiring the user to disclose to any party their ad interactions. 
    \item \cm is able to correctly apply the pricing policy of each ad without disclosing any information regarding the ad policies to users or potential competitors of the advertisers.
\end{enumerate}

%% file: Sections/04b_themis.tex
\subsection{\THEMIS: A Decentralized Ad Platform}
\label{sec:demis} 
In Section~\ref{sec:strawman}, we presented the core functionality of our ad-platform. We described (i) how the users get rewarded for their ad interactions, and (ii) how the advertisers can access the performance report of their ad campaigns.

However, the centralization and the lack of auditability of straw-man \THEMIS creates significant limitations with respect to the goals and threat model described in Section~\ref{sec:motivation}:
\begin{itemize}
[%
  \setlength{\labelwidth}{\widthof{\textbullet}}%
  \setlength{\labelsep}{13pt}%
  \setlength{\IEEElabelindent}{0pt}%
  \IEEEiedlabeljustifyl
]
  \item Advertisers need to \emph{blindly trust} the \cm with the full custody of the rewards budget set for each ad campaign.
  \item Users and advertisers \emph{have to trust} that the \cm respects the agreed policies during payouts and transfers the correct amount of rewards (step~3c in Figure~\ref{fig:diagram-strawman}).
  \item Advertisers \emph{do not} receive performance analytics of their ads (e.g. how many times an ad was viewed/clicked). Moreover, not even the \cm is able to retrieve such information.
\end{itemize}

As a result, similarly to existing approaches~\cite{privad,toubiana2010adnostic}, the entire protocol relies on the trustworthiness of a the single central authority. Moreover, users and advertisers do not have any mechanism to verify that the protocol runs as expected.

To address these issues, \THEMIS leverages a distributed \PoA ledger where business and payment logic are orchestrated by smart contracts. All participants of \THEMIS can verify that everyone runs the protocol correctly, thus requiring zero trust from any player regarding verifiability. In particular, we define two smart contracts (See Appendix~\ref{sec:appendix-smart-contracts} for full details of the smart contracts structure): 
\begin {enumerate}[A.]
    \item \emph{The  Policy Smart Contract}~(\PSC), which is responsible for the billing of users' rewards and validating the payment requests. Furthermore, it is in this smart contract that $\encpolicy$ is stored. 
    \item \emph{The Fund Smart Contract}~(\FSC), which receives and escrows the funds needed to run the campaign. The \FSC is responsible for releasing (i) the funds needed for the user rewards, (ii) the advertiser refunds, and (iii) the processing fees for the \cf. 
\end {enumerate}
In \THEMIS, instead of the central trusted authority of the \cm, we introduce the role of a 
\emph{Campaigns Facilitator}~(\cf). The responsibilities of the \cf are to 
(i)~negotiate the policies (\eg rewards per ad, impressions per ad) of the advertisers;
(ii)~deploy smart contracts in the \PoA ledger; and, lastly, (iii)~handle the on-chain payments. Our system ensures that everybody can audit and verify the behaviour of the different \cfs, so advertisers can pick the \cf they prefer to collaborate with based on their reputation. The \cf is incentivized to perform the tasks required to facilitate the ad catalog, by receiving \emph{processing fees} from advertisers.

Finally, to provide advertisers reports of performance of their ad campaigns, \THEMIS incentivises users to perform a multiparty protocol to compute ad interaction analytics in a privacy preserving manner. These participating users are referrer to as the \emph{Consensus Pool}.


\begin{figure}[t]
    \includegraphics[width=1.07\columnwidth]{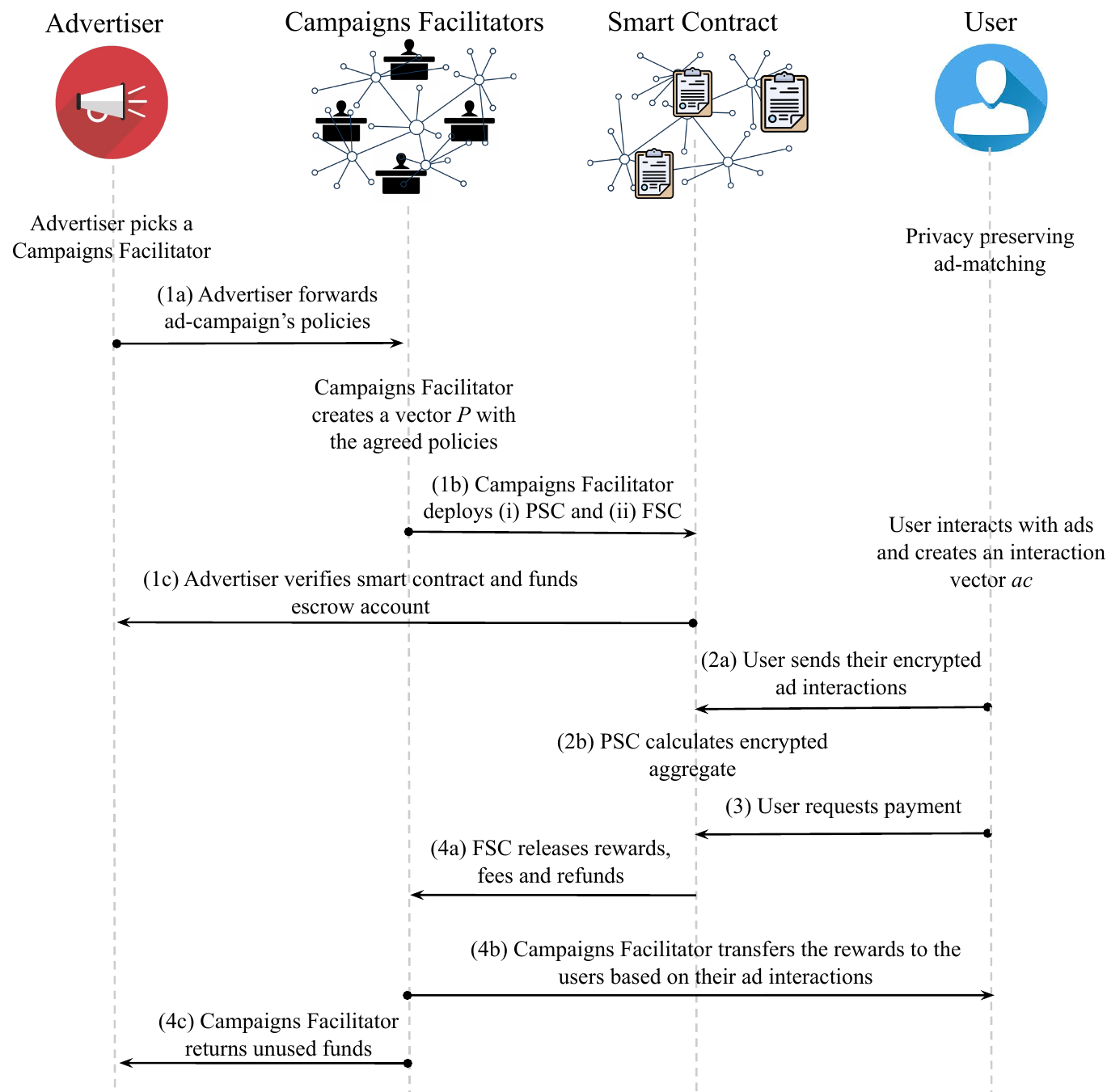}
    \caption{High-level overview of the user rewards claiming procedure in \THEMIS. This operation of \THEMIS consists of~4 different phases: (1)~Definition of Ad Rewards and set policies, (2)~Ad Reward claiming, (3)~Payment Request, (4)~Settlement of user payments and advertiser refunds.}
    \label{fig:diagram-demis}
\end{figure}

\begin{phase}{\textbf{Defining ad rewards:}}
\label{phase:def-ads-themis}
In Figure~\ref{fig:diagram-demis}, we present a high-level overview of the reward claiming procedure of \THEMIS. Similar to the straw-man approach presented in Section~\ref{sec:straw_rewards}, in order for an advertiser to include their ad campaign in the next ad-catalog facilitated by the \cf of their preference, they need to transmit their policies (\eg reward per ad) to the \cf (step 1a in Figure~\ref{fig:diagram-demis}). In order to achieve that, each advertiser exchanges a symmetric key\footnote{For the creation of this key, they follow the Diffie-Hellman key exchange protocol~\cite{DHKE}} for each ad campaign $S_i$ with the \cf. 
It then encrypts the corresponding ad campaign and sends it together with their ad creatives to the \cf. On their end, the \cf 
(i)~decrypt and check the policies are as agreed; (ii)~merge the encrypted policies of the different advertisers into the encrypted policy vector, $\encpolicy$;  and 
(iii)~deploy the two public smart contracts for this ad-catalog version (step~1b in Figure~\ref{fig:diagram-demis}).
In addition, the \cf (iv) creates a vector $\seckey$ with all the advertisers' secret keys $S_i$: 
\begin{equation}
    \seckey = \left[\seckey_1, \seckey_2 , \ldots, \seckey_{\nrads} \right] 
\end{equation}
and (v) generates a vector, $\encseckey$, that includes each of the elements in $\seckey$, encrypted with the public key ($\valikey$) of the \PoA validator nodes: 
\begin{equation}
    \encseckey =  \left[ \code{Enc}(\valikey, \seckey_1), \ldots, \code{Enc}(\valikey, \seckey_{\nrads}) \right]
\end{equation} 
Then, the \cf (vi) stores $\encseckey$ in \PSC to allow the  \PoA validators to decrypt and apply the corresponding policies on users ad interaction vectors. The process of encrypting the policy vector with symmetric keys, and then encrypting the symmetric keys with the validators public key must be seen as Hybrid Public Key Encryption~\cite{hpke}.

Once \PSC is deployed, the advertisers must verify if $\encpolicy$ encodes the policies agreed with \cf (step 1c in Figure~\ref{fig:diagram-demis}). More specifically:
\begin{enumerate}[A.]
    \item First, the advertisers fetch $\encpolicy$ vector from the public storage of the  \PSC and decrypts the policy, $\encpolicy[i]$, using the corresponding symmetric key, $\seckey_i$, and verifies it is the agreed value.

    \item Second, they fetch the escrow account address from \FSC and transfer funds to the escrow account. The amount of funds needed is determined by the number of impressions they want per ad, its part of the agreed policy $p_i$, and the processing fees to pay \cf. Once the campaign is over, the advertisers may get a refund based to the final number of impressions viewed/clicked by users. By staking the campaign's funds, the advertiser is implicitly validating the deployed ad policies.
\end{enumerate}
Once the \FSC verifies that advertisers have transferred to the escrow account the correct amount of funds, the campaign is initialized and verified. 
\end{phase}
\begin{phase}{\textbf{Claiming ad rewards:}}
\label{phase:claimrewards-themis}
Similar to what illustrated in Section~\ref{sec:strawman}, in order to claim their ad rewards, each user creates an ephemeral key pair $(\pk, \sk)$
and obtains the public threshold key $\distpk$ generated by the consensus pool (in Section~\ref{sec:aggrcount}, we describe in detail how the consensus pool is selected). By using these two keys, each user encrypts their ad interaction vector to generate two ciphertexts: 
(a)~the $\encvec$ that is used to claim ad rewards and (b)~the \encvecpublic that is used for the advertisers reporting. 

Contrary to our centralized straw-man approach, in \THEMIS, the aggregate calculation is performed via a \PSC (as can be seen in step~2b in Figure~\ref{fig:diagram-demis}). Thus, the user calls a public endpoint on \PSC and transmits both ciphertexts. To calculate the encrypted sum of the rewards the user can claim (step~2b in Figure~\ref{fig:diagram-demis}), a \PoA validator runs \PSC as follows: \begin{enumerate}
\item It decrypts each policy $\policy[i]$ using $\encseckey$ (here, \THEMIS leverages the private input transactions~\ref{sec:private_transactions}). 
\item It applies on $\encvec$ ciphertext the additively-homomorphic property of the underlying encryption scheme (as shown in Equation~\ref{eq:aggr_payout})
\item It stores the result (\ie $\encryptedaggregatedresult$) in the smart contract public store\footnote{Given that the user's public key was used for encrypting $\encvec$ ciphertext, only them can decrypt $\encryptedaggregatedresult$ and retrieve $\adclicks$.}.
\end{enumerate}
\end{phase}

\begin{phase}{\textbf{Payment request:}}
\label{phase:paymentrequest-themis}
Once the \PSC calculated the aggregate result (step~3 in Figure~\ref{fig:diagram-demis}), the user generates a payment request, $\payreq$, that, if valid, is published in \FSC. More technically, the user 
(i)~creates an ephemeral blockchain account (used only once per request) with address \addr and then (ii)~fetches and decrypts $\encryptedaggregatedresult$ to get the decrypted reward, \decryptedaggr, and generates the proof of correct decryption, $\aggrproofdec$. This way, the user 
(iii)~generates the payment request which consists of the following~3-tuple: 
\begin{equation}
    \payreq = \left[\decryptedaggr,\aggrproofdec, \addr \right].
\end{equation}
Then, (iv) the user calls a public endpoint on \PSC with the $\payreq$ encrypted with the validators keys, $\enc(\valikey, \payreq)$ as input. The function then fetches the user's aggregate, $\encryptedaggregatedresult$, decrypts the request, and verifies the zero knowledge proof, $\aggrproofdec$. If the proof is valid, it stores $\addr$ in the $\FSC$ together with the amount to be payer to this address. This way, the $\FSC$ keeps a list of buffered user payments until marked as paid. 

\end{phase}
%
%
%
\begin{phase}{\textbf{Payment settlement:}}
\label{phase:payment-settlement-themis}
The final step of the protocol regards the settlement of the user payment and advertiser refund. 

Specifically, the settlement of the user rewards in \THEMIS needs to happen in a confidential way to preserve the privacy of the total of earned rewards. To achieve this, \cf fetches the pending payments requests from \FSC, and calculates the total amount of funds required to settle all pending payments. 

As a next step (step~4a in Figure~\ref{fig:diagram-demis}), the \cf calls a public function of \FSC requesting to transfer (to an operational account owned by \cf) a given amount of tokens needed to cover the payments. If the \cf misbehaves (by requesting an incorrect amount of tokens), it will be detected, and either advertisers or users will be able to prove its misbehaviour.

Finally, \cf settles each of the pending reward payments by using a confidential payment scheme~\ref{sec:cryptoTools}. After finalizing the payments correctly (and if there are no complaints form either users or advertisers), the \cf receives from \FSC the processing fees.

In case of unused staked funds, the advertisers need to be refunded. To achieve this (step~4c in Figure~\ref{fig:diagram-demis}), \FSC utilizes the aggregate clicks per ad vector that the consensus pool has computed during the advertisers reporting (see Section~\ref{sec:aggrcount}). Based on this vector and the agreed rewards, the \FSC proceeds with returning to the advertisers the unused funds.
\end{phase}

\subsection{Privacy-Preserving Performance Analytics}
\label{sec:aggrcount}

Apart from providing users the incentives to interact with ads, in order to make an ad-platform practical, the advertisers must be able to receive feedback about their ad campaigns performance. Moreover, advertisers need to verify that the funds charged are in line with the number of times an ad was viewed/clicked by the users. Based on these statistics, advertisers get charged depending on the times their corresponding ad was clicked throughout the campaign. 

To achieve this, whenever a new version of the ad-catalog is online and retrieved from the users, a new threshold public key, $\distpk$, is generated. In order to generate such a key, a pool of multiple participating users, namely \emph{consensus pool}, is created. 
To avoid cases where there are not enough participants available online, in \THEMIS the participation in the consensus pool is incentivised\footnote{Users are incentivized to participate in this pool. Details on how to orchestrate the incentives are left outside the scope of this paper.}.
It consists of a number of selected users that have opted-in, and a smart contract responsible of the registration process and defining this pool. Any user can opt-in in the draw to become a consensus pool participant, and a random subset of all participating users is selected. 
Specifically, the smart contract keeps a time interval during which users who want to opt-in can register as participants for the draft. After that,
the smart contract utilizes an external oracle to select a random seed,
$\vrfseed$, which is used to generate random numbers.

Every registered user generates an ephemeral Verifiable Random Function key-pair,
    $(\vrfpk, \vrfsk)$,
and publishes the public key in the smart contract's public store.
Once the registration phase is closed, the smart contract calculates a threshold, $\vrfmax$, that will define the selected users
\begin{equation*}
    \vrfmax = \lfloor \frac{\nrcp}{\dpoolsize}\rfloor * p
\end{equation*}
where \dpoolsize is the size of the drawing pool (formed by all opted-in users), $\nrcp$ the expected number of participants in the distributed key generation, and $p$ an integer such that $\mathbf{Z}_p$ is the space of random numbers outputted by $\vrfrandgen(\cdot, \cdot)$.

Next, the participating users calculate their corresponding random number and a proof of correct generation using $\vrfseed$:
\begin{equation*}
    (\vrfrandom, \vrfproof) = \vrfrandgen(\vrfsk, \vrfseed).
\end{equation*}
All participants with $\vrfrandom < \vrfmax$ have won the draw, and proceed to publish $(\vrfrandom, \vrfproof)$ in the smart contract. 
Next, the selected users, namely the \emph{consensus pool}, run a distributed key generation (DKG) algorithm as defined in~\cite{dkg}. The result of running the DKG protocol is a consensus over the public key to use in the rest of the process, $\distpk$. In addition, each consensus pool participant owns a private key share $\distski i$. The distributed public key, $\distpk$, is published in \PSC to make it accessible to all users.

This key is used to encrypt a copy of the \adclicks vector (Step 2a in Figure~\ref{fig:diagram-strawman}). Hence, in addition to the \encvec illustrated in Equation~(\ref{eq:ac}), each user also sends \encvecpublic to the \PSC, where:
\begin{equation*}
\encvecpublic = \left[\ecenc(\distpk, \nrad_1), \ldots, \ecenc(\distpk, \nrad_{\nrads}) \right]
\end{equation*}
At the end of the ad campaign, the consensus pool generates the analytics report for the advertisers, showing how many times users interacted with each specific ad during the campaign. In order to generate the report, the consensus pool merges the reported \encvecpublic of every user into a single vector. This vector consists of the total number of interactions that each ad received by all users during the campaign. In order to merge all \encvecpublic of the campaign, the consensus pool performs the homomorphic addition of all reported encrypted vectors. This is possible due to the fact that every user used the same key for the encryption:
\begin{equation}
\label{eq:public-addition}
    \aggrclicks = \sum_{i\in U} \encvec^\prime_i
\end{equation}
where $U$ is the set of users and $\encvec^\prime_i$ is the vector corresponding to user $i$.  
Then, the consensus pool proceeds with the decryption, and the proof of correctness. More precisely, each consensus participant $i$ partially decrypts $\aggrclicks$: 
\begin{equation*}
    \decsharei i = \partdec(\distski i, \encryptedaggregatedclicks)
\end{equation*}
and proves it did so correctly. Finally, they post the encrypted aggregates of ads, the decrypted shares, and the proofs, to the \FSC. 

As soon as the \FSC receives at least the threshold $\threshold$ of such tuples, it combines the partial decryptions to compute the full decryption of the aggregates. This allows the advertisers to verify that the protocol ran successfully. 

First, advertisers check that the encrypted aggregates, as posted by the consensus pool, does correspond with the addition of all encryptions submitted by users. To do this, each advertiser performs the homomorphic addition as described in Equation~(\ref{eq:public-addition}), but only with their respective ads, and verify it equals the posted ciphertext. Next, they verify the proofs of correct decryption for each of the received shares. Then, they fetch the full decryption of the aggregate, representing the number of clicks/interactions their ads received. Finally, advertisers can verify that the refund received by the \FSC does correspond to the number of staked funds minus the number of views/clicks.

%% file: Sections/06_evaluation.tex
\section{Evaluation}
\label{sec:results}
In this section, we study the performance and scalability of our system. First we set out to explore the execution time of the \mbox{client-side}  operations in \THEMIS: (i) rewards claiming and (ii) payment claiming.
Then, we study the \mbox{end-to-end} execution time when multiple participants request payments in \THEMIS. Finally, we measure the overall scalability of our system and specifically, how many concurrent users claiming rewards it supports, on both single and  multi-\sidechain setups. 

\input{Sections/06a_experimental_setup}
\begin{figure}[t]
    \centering\small
    \begin{tabular}{ccc}
        \toprule
        \bf Ad Catalog & \bf Interaction & \bf Request  \\
        \bf Size (ads) & \bf Encryption (sec) & \bf Generation (sec) \\
        \midrule
        64 & 0.027 & 0.136 \\
        128 & 0.054 & 0.303 \\
        256 & 0.105 & 0.706 \\
        \bottomrule
    \end{tabular}
    \caption{Execution time of the client-side operations during reward claiming for different ad catalogs sizes: 64, 128, and 256. For an ad catalog with~256 ads it takes 0.1~ms when request generation takes around~0.7 sec.}
    \label{fig:client-side}
\end{figure}
\subsection{System performance}
\label{sec:system-performance}
In multi-client services like ad platforms, time matters for the user experience. Therefore, as a first step, we set out to explore the execution time of the requests a client issues in \THEMIS (\ie rewards claiming and payment requests) and the time it takes for the \cf to process these requests. Then, we measure the overall end-to-end time it takes for a reward request to be processed in our system.


In the case of users, we measure the time it takes for a client to generate locally a rewards claiming request for three different ad catalog sizes (64, 128 and 256 ads) and in Figure~\ref{fig:client-side} we present the results. As described in Phase~\ref{phase:claimrewards-themis}, this operation includes: 
\begin{enumerate} [(i)]
\item Interaction encryption: includes the encryption of the interaction array of the user, and
\item Request generation: includes decryption of the payment aggregate, generation of the proof of correct decryption and recovery of plaintext.
\end{enumerate}

As can be seen, the execution time to encrypt user interactions for an ad catalog of ~256 ads is as low as 0.1~sec. Similarly, for the same ad catalog size, the request generation procedure takes around~0.7~sec, proving that the client computations for reward claiming can be done on a commodity laptop or mobile device, without significant impact on the user experience. 

Apart from issuing reward claiming requests, a client also performs periodic rewards payment requests. However, such requests take place in relatively long intervals (\eg monthly) and therefore the latency imposed to the user is practically negligible.

\begin{figure}[t]
    \centering\small
    \begin{tabular}{ccc}
    \toprule
        \bf Batched & \bf Proof & \bf Verification (sec) \\
        \bf proofs & \bf generation (sec) & \\
    \midrule
        80  & 3,9 & 3 \\
        200 & 11,6 & 3 \\
        400 & 22,08 & 3 \\
        800 & 40,7 & 3\\
        \bottomrule
    \end{tabular}
    \caption{Execution time of proof generation for batches of 80, 200, 400, and 800 concurrent payments of \cf. Verification time remains constant due to the batching property of AZTEC payments.}
    \label{fig:aztec-scalability}
\end{figure}

For the settlement of the rewards payment requests (Phase~\ref{phase:payment-settlement-themis}), the \cf relies on a confidential transaction protocol to ensure the confidentiality and integrity of the payments. While implementing our system, we compared the performance of the two most popular such protocols in the context of \THEMIS: AZTEC~\cite{aztecpaper} and Zether~\cite{zether}. The conclusion is that AZTEC outperforms Zether by enabling the batching of payments into a single \mbox{on-chain} proof verification of 3 seconds. This makes the verification time constant, independently of the number of batched payments. Our results, presented in Figure~\ref{fig:aztec-scalability}, show that the \cf can achieve around~\paymentsPerDay payments/day for a batch size of~\batchsize proof payments. This results in a total of~\paymentsPerMonth payments.

As a next step, we measure the time required for the \sidechain to process concurrent payment requests end-to-end. This includes the generation of the payment requests by the user, the network latency in the communication between the clients and the \sidechain, and the time is takes for the \sidechain to process the requests and mine the blocks.

In Figure~\ref{fig:themis-several-users}, we show  the results with respect to the different concurrent users claiming their rewards (10, 30, 60, 100 users) and different ad catalog sizes (yellow bar: 64 ads, blue bar: 128 ads, green bar: 256 ads). In red, we show the time it takes for the user to decrypt the aggregate, perform the plaintext recovery locally and submit the decrypted aggregate and proof of correct decryption. As we can see,  in total for a user to claim and retrieve rewards even in the case of 100 concurrent users and a large ad catalog size of 256 ads\footnote{An ad catalog large enough handle all different ads delivered simultaneously in production systems currently in use (\ie Brave Ads~\cite{BRAVE})}:
\begin{enumerate}[A.]
	\item It takes roughly \rewardrequest for the client to request the reward calculation and retrieve the encrypted aggregate from the smart contract.
	\item It takes an additional 0.35 sec for the client to decrypt the aggregate, perform the plaintext recovery locally and submit the decrypted aggregate and proof of correct decryption.
\end{enumerate}
\noindent So it takes an overall of \rewardverif for \THEMIS to process and verify up to~100 concurrent reward payment requests of users.
 
\subsection{System Scalability}
\label{sec:system-scalability}
One of the most important challenges of \mbox{privacy-preserving} ad platforms is scalability. In the case of \THEMIS this related to how easy the system can scale with the increasing number of clients that simultaneously claim their rewards.
\input{Figs/fig-several-users}
%
%
%

In Figure~\ref{fig:themis-several-users} we saw that for an ad catalog of 256 ads and 100 concurrent users performing a payment request, it takes around~5 seconds to complete 100 concurrent payment requests. This means, that under the same conditions, the \sidechain can process around~1.7M concurrent payment requests per day, which translates to a total of \usersScalability users per month\footnote{The specifications used by the validator nodes to achieve this throughput are outlined in Section~\ref{sec:experimental-setup}.}.

\point{Horizontal scaling}
The computations performed in smart contracts in the experiments above, are highly parallelizable. However, the one-threaded event loop of the Ethereum Virtual Machine (EVM)\footnote{The EVM is the \mbox{run-time} virtual machine where the smart contract instructions are executed in each of the \mbox{validator's} machines.} does not support parallel and concurrent computations. Therefore, the EVM \mbox{run-time} becomes the scalability bottleneck when it has to handle more than 100 concurrent user requests.

To overcome this shortcoming, we tested \THEMIS  on top of multiple parallel \sidechains. Each \sidechain is responsible for one or more ad catalogs each. Although this could increase coordination complexity, the scalability gains are considerable as the number of ads processed grows linearly with the number of \sidechains. Thus, \THEMIS can scale to support millions of concurrent users per day. 
 
\input{Figs/fig-multi-sidechains}
In order to explore how horizontal scaling performs in \THEMIS, we started two and three parallel \sidechains, each using the same settings as outlined in Section~\ref{sec:experimental-setup}. Note that, by running multiple parallel \sidechains, we do not require additional or different validators. Instead, each validator is required to run three nodes, each node part of one single \sidechain.

Figure~\ref{fig:multiple-side-chains} shows the number of users \THEMIS can process by running on multiple \sidechains. As seen, the number of users increases linearly with the number of \sidechains. Assuming a setup with three parallel \sidechains, \THEMIS can handle a total of~\usersScalabilityMulti users per month~(5.1M users a day).

\subsection{Summary}
In summary, evaluation results show that \THEMIS scales linearly and seamlessly support user bases of existing centralised systems currently in production~\cite{BRAVE}. Specifically, our system
can support payment requests of around~\usersScalability  users on a single \sidechain setup or ~\usersScalabilityMulti  users on a parallel 3-\sidechain setup. 

In addition, we see that the latency users need to sustain while using \THEMIS is negligible (less that~1~sec per request payment on commodity hardware) and we show that both the \cf and \sidechain validators can also rely on commodity hardware to participate in the network.

%% file: Sections/06a_experimental_setup.tex
\subsection{Experimental Setup}
\label{sec:experimental-setup}

To study the performance and scalability of \THEMIS, we run both \mbox{client-side} and \mbox{end-to-end} measurements using the \THEMIS prototype. In this section we outline specifications over which we run our experiments. 

\point{Client specifications} The \mbox{client-side} experiments were performed on a commodity device. The device is a MacBook Pro Catalina~10.15.5, running a~2.4GHz \mbox{Qual-Core} Intel Core i5 with~16GB LPDDR3 memory.

\point{Campaign facilitator specifications} 
To study the resources necessary for campaign facilitators to participate in the network, we measured the resource and time overhead required to generate and prove the correctness of confidential payments (Phase~\ref{phase:payment-settlement-themis} of \THEMIS). In order to do so, we deployed an AWS ECS \texttt{t2.2xlarge} instance (8~vCPUs, 32~GB~RAM). 

\point{Sidechain deployment} In order to measure the performance and scalability of the \sidechain in the context of \THEMIS, we used the Mjölnir tool~\cite{mjolnir} to deploy a Quorum~\cite{quorum} \sidechain in a \mbox{production-like} environment. We deployed a 4x~Quorum \sidechain on AWS, each node running on an AWS EC2 \texttt{t2.xlarge} instance (4~vCPUs, 16~GB~RAM). All nodes are deployed in the same AWS region and part of the same subnet. For the purpose of the measurements, the network communication is considered negligible. This setup can be easily reproduced in production by setting up peering connections among different AWS Virtual Private Clouds for each of the validator organizations. The consensus protocol used by the \sidechain is the Istanbul Byzantine Fault Tolerant (IBFT) consensus protocol~\cite{EIP650}.

\point{Concurrent users} A \mbox{production-like} environment requires multiple clients requesting rewards from the \sidechain. In order to reproduce such environment, we deployed several AWS EC2 \texttt{t2.large} instances (2~vCPUs, 8~GB~RAM). We performed measurements by running 10, 30, 60, and 100 concurrent clients which request rewards from the \sidechain at roughly the same time. Using this setup, we measure the time it takes for individual clients to complete the reward calculation. In addition, we measure the \mbox{end-to-end} performance of the protocol, which includes the proof generation and verification. 

%% file: Figs/fig-several-users.tex
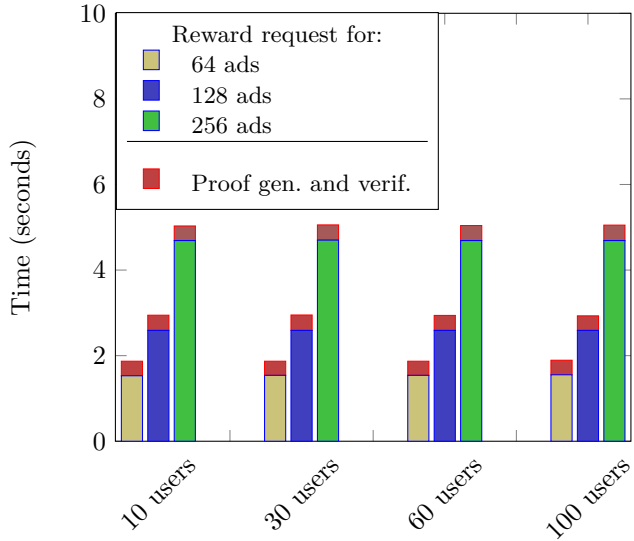
\begin{figure}[t]
    \centering
\begin{tikzpicture}[
  every axis/.style={ 
    ybar stacked,
    ymin=0,ymax=10,
    ylabel=Time (seconds),
    x tick label style={rotate=45},
    symbolic x coords={
      10 users, ,
      30 users, ,
      60 users, ,
      100 users, 
    },
  bar width=8pt
  },
]

\begin{axis}[bar shift=-10pt,name=boundary]
\addplot+[fill=yellow!50!gray] coordinates
{(10 users,1.53) (30 users,1.54) (60 users,1.54) (100 users, 1.553)};\label{64}
\addplot+[fill=red!50!gray] coordinates
{(10 users,0.34) (30 users,0.33) (60 users,0.33) (100 users, 0.34)};\label{overhead}
\end{axis}

\begin{axis}[hide axis]
\addplot+[fill=blue!50!gray] coordinates
{(10 users,2.59) (30 users,2.59) (60 users,2.59) (100 users, 2.59)};\label{128}
\addplot+[fill=red!50!gray] coordinates
{(10 users,0.355) (30 users,0.36) (60 users,0.35) (100 users, 0.34)};
\end{axis}

\begin{axis}[bar shift=10pt,hide axis]
\addplot+[fill=green!50!gray] coordinates
{(10 users,4.69) (30 users,4.7) (60 users,4.69) (100 users, 4.69)};\label{256}
\addplot+[fill=red!30!gray] coordinates
{(10 users,0.34) (30 users,0.355) (60 users,0.35) (100 users, 0.36)};
\end{axis}

\node[draw,fill=white,inner sep=4pt,above=-1.3cm, left=-4.3cm] at (boundary.north west) {\small
    \begin{tabular}{cl}
    \multicolumn{2}{c}{Reward request for:}\\
    \ref{64} & $64$ ads\\
    \ref{128} & $128$ ads\\
    \ref{256} & $256$ ads \\
    \hline \\
    \ref{overhead} & Proof gen. and verif.
    \end{tabular}};
\end{tikzpicture}

    \caption{Cumulative time (in seconds) for the \THEMIS protocol runs with~64 (yellow),~128 (blue), and~256 (green) ads. In red, the overhead caused by proving and verifying correct decryption. Results presented for number of concurrent clients requesting the reward computation from the smart contract.}
    \label{fig:themis-several-users}
\end{figure}

%% file: Figs/fig-multi-sidechains.tex
\begin{figure}[t]
    \centering
\begin{tikzpicture}[
  every axis/.style={ 
    ybar stacked,
    ymin=0,ymax=160,
    ylabel=Millions Users per Month,
    xlabel=Number of Sidechains,
    symbolic x coords={1, ,2, ,3},
  bar width=8pt
  },
]

\begin{axis}[name=boundary]
\addplot+[fill=blue!50!gray] coordinates
{(1,51) (2,100) (3,153)};
\end{axis}

\end{tikzpicture}

    \caption{Number of Millions of Users that \THEMIS can handle per Month by deploying \sidechains. The expected linear growth is confirmed by our experiments.}
    \label{fig:multiple-side-chains}
\end{figure}
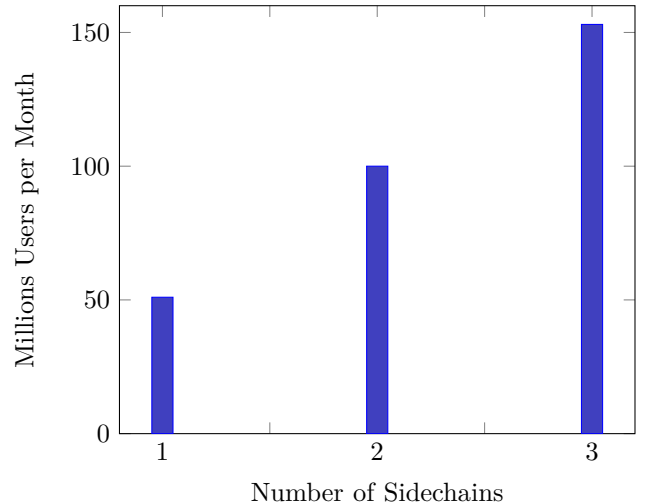

%% file: Sections/07_discussion.tex
\section{Discussion}
\label{sec:misbehavingCF}

In \THEMIS a \cf can cheat in two ways: 
(1)~as it is the entity orchestrating the confidential payments, it may send incorrect rewards to users or, 
(2)~it could use its power to send rewards not only to the user but to other accounts of their control. Both of these actions may be discovered by either users or advertisers.
\begin{enumerate}[A.]
    \item In case of scenario~(1) The users can provably challenge \cf for incorrect behaviour by proving that the payment received does not correspond to the payment request they generated. To do so, the user calls the \FSC to prove that the amounts received by the private payment does not correspond to the decrypted aggregate in the payment request $\payreq$. We stress that in case a user must undergo such a scenario, only the aggregate amount of a single ad-catalog will be disclosed (and not its interaction with ads). 
    \item In case of scenario~(2), the escrow account will not have enough funds, resulting in some advertiser getting a smaller refund to what is stated in the performance report of their ad campaign (as described in Section~\ref{sec:aggrcount}). 
    In this case, the advertiser can prove that the received refund does not correspond to the amount staked in Phase~\ref{phase:def-ads-themis} of \THEMIS, minus the rewards paid to users based on the numbers of clicks their ads received.
\end{enumerate}
To claim misbehaviour, users and advertisers can file a complaint via a public function on \FSC (that validates the complaint). If any  complaints are filed, the \FSC switches its state to ``failed'' and \cf will not receive any processing fees, something that affects their reputation.

%% file: Sections/08_related.tex
\section{Related Work}
\label{sec:related}
The current advertising ecosystem abounds with issues associated with its performance, its transparency, the user's privacy and the integrity of billing and reporting. These failures are already well studied and there are numerous works aiming to shed light on how digital advertising works~\cite{goldstein2013cost,vallina2012breaking,pachilakis2019no,gill2013best,reznichenko2011auctions,rtbPrices17,bashir2018diffusion}.

Apart from the studies highlighting the failures of current ad delivery protocols 
there are also important novel ad systems proposed. 
In~\cite{10.5555/646139.680791}, Juels is the first to study private targeted 
advertising. Author proposes a privacy-preserving targeted ad delivery scheme 
based on PIR and Mixnets. In this scheme, advertisers choose a negotiant function that assigns the most fitting ads in their database for each type of profile.  The proposed scheme relies on heavy cryptographic operations and therefore it suffers from intensive computation cost. Their approach focuses on the private distribution of ads and does not take into account other aspects such as view/click reporting. 

In~\cite{toubiana2010adnostic}, authors propose Adnostic: an architecture to enable users to retrieve ads on the fly. Adnostic prefetches \emph{n} ads before the user starts browsing and stores them locally. Aside from the performance benefits of this strategy, Adnostic does this prefetching also in order to preserve the privacy of the user. The parameter \emph{n} is configurable: larger \emph{n} means better ad matching, when smaller \emph{n} means less overhead. In order for the ad-network to correctly charge the corresponding advertisers, Adnostic performs secure billing by using homomorphic encryption and zero-knowledge proofs.

In~\cite{haddadi2009not,privad,reznichenko2014private}, authors propose Privad: an online ad system that aims to be faster and more private than today's ad schema. Privad introduces an additional entity called Dealer. The Dealer is responsible for anonymizing the client so as to prevent the ad-network from identifying the client and also handle the billing. To prevent the Dealer from accessing user's behavioral profile and activity it encrypts the communications between the client and the Dealer. A limitation of Privad is that Dealer is a centralized entity that needs to be always online. 

In~\cite{backes2012obliviad}, authors propose ObliviAd: a provably secure and
practical online behavioral advertising architecture that relies on a secure remote co-processor (SC) and Oblivious RAM (ORAM) to provide the so called secure hardware-based PIR. In ObliviAd, to fetch an ad, a user first sends their encrypted  behavioral profile to the SC which securely selects the ads that match best based on the algorithm specified by the ad network. To prevent the ad-network from learning which ads are selected, they leverage an ORAM scheme. The selected ads are finally sent to the user encrypted, along with
fresh tokens used to billing. User will send back one of these tokens as soon as they view/click on an ad.

In~\cite{7164916} authors point out that, in current advertising systems
the ad-network exclusively determines the payment to get from advertisers and the revenue to share with publishers. This means that (i) a malicious ad-network can overcharge advertisers or underpay publishers.  To make matters worse, as bills cannot be justified by the ad-network, malicious advertisers
can deny actual views/clicks to ask for refund.  On the other hand, (ii) malicious publishers may claim clicks that did not happen, in order to demand higher revenues. To address this problem of unfairness, authors propose a protocol where the ad click reports are encrypted by the user using the public key of the ad-network and signed by both publishers and advertisers.

In~\cite{green2016protocol}, authors use an additively encryption scheme to design a protocol that enables privacy-preserving advertising reporting at scale, without needing any trusted hardware. Performance evaluation results show that their protocol reduces the overhead of reporting by orders of magnitude compared to the ElGamal-based solution of Adnostic~\cite{toubiana2010adnostic} (\ie  1 MB  of  bandwidth  per impression when handing 32,000 advertisements). Contrary to our approach, authors assume a Trusted Third Party (TTP) that owns the key for the homomorphic encryption.

In~\cite{10.1145/2462456.2464436}, authors propose
CAMEO: a framework for mobile advertising that employs intelligent and proactive prefetching of advertisements. CAMEO uses context prediction, to significantly reduce the bandwidth and energy overheads, and provides a negotiation protocol that empowers applications to subsidize their data traffic costs by ``bartering'' their advertisement rights for access bandwidth from mobile ISPs. 
In~\cite{10.1145/1859983.1859993}, authors propose a location-aware, personalised and private advertising system for mobile platforms. In this system, ads are locally broadcast to users within mobile cells. The ad matching happens locally based on the user interests. Finally ad view and click reports are collected using a DTN system. In~\cite{biswas2014privacy}, authors propose a new ad protocol that uses homomorphic and searchable encryption  to allow users transmit mobile sensor data to a cloud service that responds back with the best matching contextual advertisements. 

In~\cite{10.1145/2534169.2486038}, authors present VEX, a protocol for ad exchanges to run low-latency and high-frequency ad auctions that are verifiable and auditable, in order to prevent fraud in a context where parties participating in the auction -- bidders and ad exchanges -- may not know each other. Based on their evaluation of the system, the authors claim that the additional storage required and latency imposed by VEX are low and practical in the context of ad auctions. 
In~\cite{pang2015prota}, authors present and implement PROTA, a privacy-preserving protocol for \mbox{real-time} advertising which uses keywords to match users interests with ads. By using bloom filters, the authors make the ad matching task efficient. The protocol relies on a trusted third party to cooperate with the ad exchange during the bidding and ad delivering phase. The authors implement and evaluate the protocol, and conclude that the time upper bond for matching ads is 200ms, which is considerable practical in the context of an ad matching system.

In~\cite{5958026}, authors present and evaluate a system that aims at providing high-quality ad targeting in multiple scenarios, while giving the user the ability to control their privacy. The system consists of tailored extensions that \emph{mine} the user behaviour locally with low overhead. The extensions generate user behavioural data that can be shared with advertisers without leaking undesirable user information. Similarly to \THEMIS, the authors discuss how the system can be used by users and advertisers, and how it can be used as a replacement for the tracking-based business model in the online advertising industry.

In~\cite{8228673}, authors set out to formalize the concept of privacy in the context of the online advertising ecosystem and to develop a provably secure privacy-preserving protocol for the online advertising ecosystem. While the authors claim that the definition of privacy presented in the paper is more useful compared to previous work in the online advertising context, their attempts to develop a provably secure privacy-preserving protocol has failed due to being hard to balance privacy with usefulness of the user data. The authors conjecture that cryptographic mechanisms have the potential to solve the privacy versus data usefulness conundrum. Using applying cryptography is the basis of how \THEMIS proposes to preserve privacy when calculating ad rewards, providing advertisers with campaign metrics and performing confidential payments to users.

Towards a similar direction with the user rewarding schema of \THEMIS, in~\cite{wang2015privacy}, authors
propose a privacy-aware framework to promote targeted advertising. In this framework, an ad broker responsible for handling ad targeting, sits between advertisers and users and provides certain amount of compensation to incentivize users to click ads that are interesting yet sensitive to them. 
In~\cite{parra2017pay}, authors propose a targeted advertising framework which enables users to get compensated based on the amount of user tracking they sustain and the privacy they lose. The authors analyze the interaction between the different parties in the online advertising context --- advertisers, the ad broker and users --- and propose a framework where the interactions between the different parties are a positive-sum game. In this game, all parties are incentivized to behave according to what other parties expect, achieving an equilibrium where everyone benefits. More specifically, the users determine their click behaviour based on their interested and their privacy leakage, which in turn will influence the advertisers and ad broker to provide less invasive and better ads. \THEMIS relies on a similar game theoretical approach. By providing compensation for good behaviour while providing the verification mechanisms for all parties to audit whether everyone is behaving according to the protocol, the incentives to cheat and misbehave are lower.

%% file: Sections/09_conclusions.tex
\section{Conclusions}
\label{sec:conclusions}
In this paper, we presented \THEMIS, the first decentralized, scalable,  and privacy-preserving ad platform that provides integrity and auditability to its participants, so users do not need to blindly trust any of the protocol actors. To increase the user engagement with ads and provide advertisers with the necessary performance feedback about their ad campaigns, \THEMIS (i) rewards users for their ad viewing and (ii) provides advertisers with verifiable and privacy-preserving campaign performance analytics.

We implemented our approach by leveraging a permissioned blockchain with Solidity smart contracts as well as zero-knowledge techniques. We evaluate the scalability and performance of our prototype and show that \THEMIS can support reward payments for more than~\usersScalability users per month on a single-\sidechain setup and~\usersScalabilityMulti  users on a parallel multi-\sidechain setup, proving linear scalability.

While many projects and companies have proposed the use of blockchain for online advertising, we believe that \THEMIS is the first system that delivers on that promise. Given the practicality of the approach and the combination of security, privacy, and performance properties it delivers, \THEMIS can be used as a foundation of a radically new approach to online advertising. 




%% file: Sections/10_appendix_smart_contract.tex
\section{Smart Contracts}
\label{sec:appendix-smart-contracts}

\newcommand{\scfunction}[1]{\underline{\texttt{#1}}}
\newcommand{\nrpolicies}{l}

\newcommand{\fundslist}{\mathcal{F}}
\newcommand{\fundi}[1]{\fundslist_{#1}}
\newcommand{\tokensamount}{\tau}
\newcommand{\paymentrequests}{\texttt{PaymentReq}}
\newcommand{\payedrequests}{\texttt{PayedReq}}
\newcommand{\processingfees}{\omega}
\newcommand{\campaigninit}{\texttt{Init}}
\newcommand{\txref}{\texttt{TxRef}}

In this section, we specify the functionality and properties of the smart contracts necessary to run \THEMIS. In practice and at the EVM level, the policy smart contract logic and fund smart contract logic may be split into multiple smart contracts, but for the sake of simplicity, we describe the logic as part of two smart contracts per \THEMIS campaign: (i) the Policy Smart Contract (Figure~\ref{fig:psm}), and (ii) the Fund Smart Contract (Figure~\ref{fig:fsm}). We assume that both smart contracts have access to storage of the $\cf$'s public key and the consensus participants public key (when generated).

\subsection{Policy Smart Contract}
\point{Public data structures} The Policy Smart Contract (\PSC) keeps its state in three public data structures: an array with the encrypted ad policies ($\encpolicy$), an array containing all rewards aggregates (\texttt{Agg}) calculated by the smart contract, and an array of encrypted symmetric keys, $\encseckey$, used to encrypt the policies. The latter allows the validators to decrypt the policies and apply them to the aggregate requests. 

\point{$\texttt{StorePolicy()}$} This private function can be called only by the account which deployed the smart contract, \ie by the Campaign Facilitator (\cf). The function receives an array of $\texttt{uint265}$ types -- which represent the encrypted ad policies for the campaign agreed with the advertisers (Phase~\ref{phase:def-ads-themis} in Section~\ref{sec:demis}) --- and it initializes the public ($\encpolicy$) data structure with its input. Each policy is encrypted using a symmetric key agreed between the \cf and the corresponding advertiser.

\point{$\texttt{ComputeAggregate()}$} It is a public function that is exposed to the users. Users call this function with an array containing their encrypted interactions (Phase~\ref{phase:claimrewards-themis} in Section~\ref{sec:demis}). The smart contract proceeds to calculate the reward aggregate based on the user input and the ads policies ($\encpolicy$). The aggregate is stored in the (\texttt{Agg}) data structure, which is accessible to all users.

\point{$\texttt{GetAggregate()}$} It is a public function which receives an ID (\ie $\texttt{uint265}$) and returns the encrypted aggregate indexed by the respective ID in the \texttt{Agg} data structure. This function is used by users to fetch the encrypted aggregate calculated by the smart contract, after having called the $\texttt{ComputeAggregate()}$ function described above.

\point{$\texttt{PaymentRequest()}$} It is a public function which receives an encrypted payment request from users (Phase~\ref{phase:paymentrequest-themis} in Section~\ref{sec:demis}) under the validators' key. If the payment request is valid, the smart contract buffers the request in the Fund Smart Contract ($\paymentrequests$). The buffered payments are settled periodically by \cf.

\input{Sections/appendix_code/pcm}

\subsection{Fund Smart Contract}
\point{Public data structures} The Fund Smart Contract (\FSC) keeps it state in multiple public data structures. The $\campaigninit$ parameter represents whether the ad campaign has started. In order for the $\campaigninit$ to turn to be marked as initialized (\ie $true$), all campaign advertisers -- which are kept in $\advs$ -- must confirm their participation by depositing the funds (to the ad campaign's escrow account $\fundslist$) necessary to cover their ad campaign in the smart contract account. In addition, the \FSC keeps a list of all payment requests triggered by \PSC and the successfully payed requests $\payedrequests$. Finally it stores the agreed processing fees of the \cf, and a value of the overall ad interaction, $\aggrclicks$, which is updated by the consensus pool.

\point{$\texttt{StoreAdvID()}$} Private function that can be called by the \cf to add new advertisers to the campaign. If the campaign has not started (\ie $\campaigninit: false$), the advertisers ID is added to the $\advs$ list. No new advertisers can be added after a campaign has started.

\point{$\texttt{StoreAggrClicks()}$} This public function has an access control policy that only a value signed by the consensus participants will update the public data structure. It is a function that is used to update the state of the smart contract overall ad interactions, $\aggrclicks$.

\point{$\texttt{StoreFunds()}$} Public function called by the advertisers upon transferring the campaign funds to the \FSC account. When all the advertisers have transferred the funds necessary to cover their ad campaign, the smart contract updates its state to $\campaigninit: true$.

\point{$\texttt{InitialiseCampaign()}$} Private function that is only called by the smart contract. It sets $\campaigninit: true$.

\point{$\texttt{SettlementRequest()}$} This public function has an access control policy that only the \cf can successfully call it. It requests the release of funds to the \cf account, in order for the \cf to be able to have the funds for settling the buffered payment requests (in \PSC).

\input{Sections/appendix_code/fcm}

\point{$\texttt{RefundAdvertisers()}$} This function is called internally by the \FSC and it is triggered either when: (1) all the ads in the campaign have been "spent" by the users or (2) when a pre-defined \textit{epoch} has passed, signaling the end of the campaign. This function releases the funds to the advertisers, based on the $\aggrclicks$ per advertiser in the campaign.

\point{$\texttt{PayProcessingFees()}$} This function is -- alongside with $\texttt{RefundAdvertisers()}$ -- called when the campaign is finished. It verifies if the \cf has behaved correctly and pays the fees to the \cf's account.

\point{$\texttt{RaiseComplaint()}$} This public function allows users to prove that the \cf misbehaved. In order to cull such a function, users must prove that their corresponding aggregate does not correspond to the private payment they received. To this end they disclose the aggregate value they were expecting to earn. This will prove that the \cf misbehaved and the smart contract can flag the latter as such.

\point{$\texttt{ClaimInsufficientRefund()}$} This public function allows advertisers to prove they have received insufficient refunds. To this end, advertisers simply call this function which automatically checks the validity of the claim. If the claim is valid, it flags the \cf as misbehaving.

%% file: Sections/appendix_code/pcm.tex
\begin{figure}[t]
\begin{cvbox}[Policy Smart Contract]
\scfunction{Public Storage}
\begin{itemize}
    \item Enrypted agreed policies $\encpolicy: \left[\right]$
    \item Aggregates \texttt{Agg}: $\left[\right]$
    \item Encrypted keys used to encrypt the ad policies: $\encseckey:\left[\right]$
\end{itemize}
\vspace{2mm}
\scfunction{StorePolicy}

Inputs: Encrypted policy, $\encpolicy_i$.
\begin{enumerate}
    \item Set $\encpolicy\left[i\right] = \encpolicy_i$
\end{enumerate}
\vspace{2mm}
\scfunction{StoreEncryptedKeys}

\noindent Inputs: Encrypted keys, $\encseckey*$, and signature, $\signature$ of the call
\begin{enumerate}[leftmargin=*]
    \item Verify $\signature$ with the \cf's public key, $\cfpk$.
    \item Set $\encseckey=\encseckey*$
\end{enumerate}

\underline{\texttt{ComputeAggregate}}

\noindent Inputs: the user's public key $\pk_{user}$, and the 
encrypted ad interaction, $\encryptedad$.
\begin{enumerate}
    \item Require 
    $\texttt{length}(\encryptedad) == \texttt{length}(\encpolicy)$
    \item Decrypt the encrypted keys 
    \[\seckey = \left[\dec\left(\sk_{\mathcal{V}},\encseckey[i]\right)\right]_{i=1}^N\] 
    \item Use these keys to decrypt all entries of $\encpolicy$
    \[ \policy = \left[\dec\left(\seckey[i],\encpolicy[i]\right)\right]_{i=1}^N\]
    \item Store aggregate
    \[\texttt{Agg}\left[\pk_{user}\right] = \sum_{i=1}^{N} \policy\left[i\right] \cdot \encvec\left[i\right]\]
\end{enumerate}
\vspace{2mm}
\scfunction{GetAggregate}

\noindent Inputs: the user's public key $\pk_{user}$
\begin{enumerate}
    \item Return $\texttt{Agg}\left[\pk_{user}\right]$
\end{enumerate}
\vspace{2mm}
\scfunction{PaymentRequest}

\noindent Inputs: The user's public key, $\pk_{user}$, decrypted aggregate, proof of decryption, and the address receiving the payment:
\[
\payreq=\left[\pk_{user},\decryptedaggr, \signaturereward, \aggrproofdec, \addr \right]
\]
encrypted under the validators key, $\enc(\valikey, \payreq)$
\begin{enumerate}
    \item Get the encrypted aggregate related to the public key, 
    \[\texttt{Agg}\left[\pk_{user}\right]\]
    \item Decrypt the encrypted payment request, to get $\payreq$
    \item Verify $\signaturereward$ with \cf's public key, $\cfpk$. If it validates, then;
    \item Verify proof of decryption $V=\aggrproofdec.\texttt{Verif}()$. If $V=\top$, append $\addr$ to the payment request $\paymentrequests$ variable of the \textbf{Fund Smart Contract}.
\end{enumerate}
\vspace{2mm}
\end{cvbox}
\caption{Description of the public storage and functionality of the Policy Smart Contract (PSC)}
\label{fig:psm}
\end{figure}

%% file: Sections/appendix_code/fcm.tex
\begin{figure*}[t]
\begin{cvbox}[Fund Smart Contract]
\begin{minipage}[t]{.45\textwidth}
\scfunction{Public Storage}
\begin{itemize}
    \item Campaign initialised, $\campaigninit: false$
    \item Advertisers list, $\advs: \left[\right]$
    \item Ad campaign's escrow account, $\fundslist: \left[\right]$
    \item Agreed processing fees, $\processingfees$
    \item Payment request, $\paymentrequests: \left[\right]$
    \item Payed requests, $\payedrequests: \left[\right]$
    \item Overall ad interactions, $\aggrclicks: 0$
\end{itemize}
\vspace{2mm}
\scfunction{StoreAdvID}

\noindent Inputs: Advertiser's id $\advs_{id}$ that has agreed an ad policy with \cf
\begin{enumerate}
    \item Append to the advertisers list, the advertiser's id 
    
    \centerline{$\advs = \advs \| \advs_{id}$}
\end{enumerate}
\scfunction{StoreAggrClicks}

\noindent Inputs: Number of ad clicks performed by all users, $\aggrclicks'$ and signature of the value $\signature$. 
\begin{enumerate}
    \item Verify $\signature$ with $\cp$ public key. If it validates, then
    \item Set $\aggrclicks = \aggrclicks + \aggrclicks'$
\end{enumerate}
\vspace{2mm}
\scfunction{StoreFunds}

\noindent Inputs: Fund for the campaign, $\fundi{id}$ which is determined by the number of impressions the advertiser wants per ad, the agreed policies and the processing fees. And the advertiser's ID, $\advs_{id}$.
\begin{enumerate}
    \item Store fund in escrow account $\fundslist\left[id\right] = \fundi{id}$
    \item If $\fundslist$ contains funds from all advertisers in $\advs$, call 
    \item[] \texttt{InitialiseCampaign}
\end{enumerate}
\vspace{2mm}
\scfunction{InitialiseCampaign}

\begin{enumerate}
    \item Set $\campaigninit = true$
\end{enumerate}
\vspace{2mm}
\scfunction{SettlementRequest}

\noindent Input: amount, $\tokensamount$, and a signature of the request, $\sigma$
\begin{enumerate}
    \item Verify $\sigma$ with \cf public key, $\cfpk$. If it validates, then;
    \item Send $\tokensamount$ to \cf's account.  
\end{enumerate}
\vspace{2mm}
\scfunction{PaymentProcessed}

\noindent Input: reference of the private transaction, $\txref$ to address $\addr$
\begin{enumerate}
    \item Check existence of $\txref$, if it exists,
    \item Add $\addr$ to $\payedrequests$
    \item If $\payedrequests == \paymentrequests$, then call 
    \item[] \texttt{PayProcessingFees}($\processingfees$) and
    \item[] \texttt{RefundAdvertisers}($\advs, \aggrclicks$)
\end{enumerate}
\vspace{2mm}
\end{minipage}
\hspace{1cm}
\begin{minipage}[t]{.45\textwidth}
\scfunction{RefundAdvertisers}

\noindent Input: list of advertisers, $\advs$, and total value of aggregate clicks, $\aggrclicks$.
\begin{enumerate}
    \item If $\fundslist\left[id\right]$ is greater than the amount "spent" during the campaign, refund the corresponding amount to $\advs_{id}$.
    \item If  $\fundslist\left[id\right]$ is smaller than the amount "spent", request payment to $\advs_{id}$.
\end{enumerate}
\vspace{2mm}
\scfunction{PayProcessingFees}

\noindent Input: agreed processing fees for the campaign, $\processingfees$, between \cf and advertisers.
\begin{enumerate}
    \item Pay $\processingfees$ to \cf's account.
\end{enumerate}
\vspace{2mm}
\scfunction{RaiseComplaint}

\noindent Input: user's public key, $\pk_{user}$, reference to private payment, $tx_{priv}$, and the opening of the latter, $(r, l)$, where $r$ is the hiding value and $l$ the payment amount.
\begin{enumerate}
\item Verify that $tx_{priv}$ corresponds to a transaction with $(r, l)$. If yes, 
\item Fetch the payment request of the user, $\paymentrequests[\pk_{user}]$
\item Decrypt the request to extract the aggregate
\item If $l$ does not correspond to the amount fetched, flag the \cf as dishonest and append the proof, i.e. the decrypted aggregate and private payment opening of the corresponding user.
\end{enumerate}  
\vspace{2mm}
\scfunction{ClaimInsufficientRefund}

\noindent Input: advertiser's id, $\advs_{id}$.
\begin{enumerate}
\item Fetch the funds submitted by the advertiser, $\fundslist[\advs_{id}]$
\item Fetch the number of views received by the ads of the advertisers, and multiply them by the agreed policy value, to compute the total amount spent by the advertiser
\[\texttt{Sp} = \sum_{i\in S}\policy[i] * \aggrclicks[i]\]
where $S$ is the set of all ads corresponding to advertiser $\advs_{id}$. 
\item If $\texttt{Sp}$ plus the refund paid to the advertiser does not correspond to $\fundslist[\advs_{id}]$, flag the \cf as dishonest.
\end{enumerate}
\end{minipage}
\end{cvbox}
\caption{Description of the public storage and functionality of the Fund Smart Contract (FSM)}
\label{fig:fsm} 
\end{figure*}